\documentclass[draft]{agujournal2019}
\usepackage{lineno}
\draftfalse

\usepackage{url}
\usepackage[finalnew]{trackchanges} 
\usepackage[utf8]{inputenc}
\usepackage{soul}
\usepackage[amsmath,upgreek]{}
\journalname{Journal of Geophysical Research: Planets}

\begin{document}
	
	
	\title{Martian Dust Storms and Gravity Waves: Disentangling Water Transport to the Upper Atmosphere}
	
	
	\authors{Dmitry S. Shaposhnikov\affil{1}, Alexander S. Medvedev\affil{2}, Alexander V. Rodin\affil{1,3}, Erdal Yi\u{g}it\affil{4}, and Paul Hartogh\affil{2}}
	
	\affiliation{1}{Moscow Institute of Physics and Technology, Moscow, Russia}
	\affiliation{2}{Max Planck Institute for Solar System Research, G\"ottingen, Germany}
	\affiliation{3}{Space Research Institute, Moscow, Russia}
	\affiliation{4}{George Mason University, Fairfax, VA, USA}
	
	\correspondingauthor{D. S. Shaposhnikov}{shaposhnikov@phystech.edu}
	
	
	\begin{keypoints}
		\item The impact of gravity waves and planet-encircling dust storms on the hydrological cycle is explored with a global circulation model
		\item Water enters the upper atmosphere mainly along upward branches of the circulation, while molecular diffusion plays little role
		\item Gravity waves modulate the onset and strength of upward water ``pump" channels during equinoctial and solstitial dust storms differently
	\end{keypoints}
	
	
	\begin{abstract}
		Simulations with the Max Planck Institute Martian general circulation model for Martian years 28 and 34 reveal details of the water ``pump" mechanism and the role of gravity wave (GW) forcing. Water is advected to the upper atmosphere mainly by upward branches of the meridional circulation: in low latitudes during equinoxes and over the south pole during solstices. Molecular diffusion plays little role in water transport in the middle atmosphere and across the mesopause. GWs modulate the circulation and temperature during global dust storms, thus changing the timing and intensity of the transport. At equinoxes, they facilitate water accumulation in the polar warming regions in the middle atmosphere followed by stronger upwelling over the equator. As equinoctial storms decay, GWs tend to accelerate the reduction of water in the thermosphere. GWs delay the onset of the transport during solstitial storms and change the globally averaged amount of water in the upper atmosphere by 10-25\%.
	\end{abstract}
	
	\section*{Plain Language Summary}
	
	Transport of water to the Martian upper atmosphere with subsequent photodissociation and escape of hydrogen into space occurs near perihelion and amplifies during global dust storms. One explanation is the so-called water ``pump" due to the seasonally varying global meridional circulation. Forcing by atmospheric gravity waves is a major mechanism that drives this circulation in the middle and upper atmosphere. The interplay between gravity waves, dust storms and transport of water from the lower atmosphere to the thermosphere has been explored with a Martian general circulation model. When dust storms take place at equinoxes, gravity waves enhance warming in polar regions and accumulation of vapor there. They also strengthen the uplift in low latitudes leading to a significant increase of water in the thermosphere. When global dust storms start to decay, wave forcing helps to remove water from upper layers of the atmosphere more abruptly. When storms onset near the perihelion solstice (southern hemisphere summer), gravity waves delay the upward transport of water over the south pole first, but then intensify it and the redistribution across the globe. Thus, more water is delivered to the thermosphere during dust storms and, potentially, \add{can} escape to space as atomic hydrogen.
	
	
	\section{Introduction}
	
	Gravity waves (GWs) are omnipresent in the atmosphere of Mars and continuously disturb it at all heights. Recent observations not only provided ample evidence for their existence, but also determined their seasonal and spatial climatology in the lower \cite{heavens2020multiannual} and upper atmosphere \cite{ jesch2019density,  Siddle_etal19, vals2019study, Leelavathi_etal20, Nakagawa_etal20b, Li_etal21, Starichenko_etal21,Yigit_etal21a, Yigit_etal21b}. The main dynamical role of GWs in planetary atmospheres is to re-distribute momentum and energy between atmospheric layers, thus providing a vertical coupling between the lower and the upper atmosphere \cite{yiugit2015internal, yiugit2019obscure}. On Mars, various phenomena associated with GWs have been studied over the recent decade \cite<e.g., see the review by>[]{medvedev2019gravity}, however among the least explored remains the aspect of how GWs influence the transport of minor species.  
	
	Water is an extremely important minor constituent in the Martian atmosphere. Although the abundance of water vapor in the atmosphere is very small (of the order of a few hundredths of a percent), it is the main source of hydrogen in the thermosphere \cite{hunten1970production, Stone_etal20}. Hydrogen is produced by photodissociation of H$_2$O molecules delivered to the upper atmosphere (above $\sim$80 km) from below and, due to its low molecular mass, has the largest probability of escaping into space compared to other species. It was suggested that hydrogen escape was the main channel of atmospheric loss and dehydration of Mars in the past \cite{mcelroy1972mars, parkinson1972spectroscopy}. There is plentiful observational evidence that the amount of hydrogen atoms near the exobase varies with seasons by an order of magnitude with the maximum near perihelion (solar longitude $L_s\approx270^\circ$) \cite{bhattacharyya2017seasonal, chaufray_etal21, halekas2017seasonal}. Seasonal variations have been found for water in the lower atmosphere \cite{smith2002annual, smith2009compact, maltagliati2011annual, maltagliati2013annual, Fedorova_etal21}. Both hydrogen in the exosphere and water in the upper atmosphere increase significantly during major dust storms \cite <e.g.,>[]{bhattacharyya2015strong, chaffin2014unexpected, clarke2014rapid, chaffin2017elevated, fedorova2018water, fedorova2020stormy, heavens2018hydrogen, aoki2019water}. The observed seasonal behavior and the dependence on atmospheric dust has been explained by the water ``pump" mechanism \cite{shaposhnikov2019seasonal}, according to which water is transported from the lower to upper atmosphere by the meridional circulation.
	Similar results for the dust storm of MY34 have been produced in the simulation study of \citeA{neary2020explanation}, which showed that the water vapor distribution at higher altitudes is sensitive to the distribution of dust. These authors emphasized the role of direct heating by airborne aerosol and resulting inhibition of ice cloud formation in increasing the amount of water in the middle atmosphere.
	
	Our study examines the water transport mechanism further by including into consideration the forcing provided by the subgrid-scale GWs, a mechanism that has not been explored in the context of water transport so far. This is done using simulations with our recently developed hydrological scheme \cite{shaposhnikov2018modeling, shaposhnikov2019seasonal} and the whole atmosphere GW parameterization \cite{Yigit_etal08} implemented in the Max Planck Institute Martian general circulation model (MPI--MGCM). Since high-altitude water abundances increase during dust storms, we selected Martian years 28 (MY28) and 34, when major planet-encircling dust events have occurred near the solstice and equinox, correspondingly. The main focus of the study is on MY34, because simulations for MY28 have been already described in some detail \cite{shaposhnikov2019seasonal}. 
	
	In section~\ref{sec:MGCM}, we outline the model and setup of numerical experiments. Then, in section~\ref{sec:storm}, we discuss the impact produced by the MY34 dust storm on the water cycle, and briefly compare model temperatures with the measurements of the Mars Climate Sounder instrument onboard Mars Reconnaissance Orbiter (MCS--MRO). The relative roles of advection and molecular diffusion are explored in section~\ref{sec:adv-diff}. Differences introduced by GWs are described in section~\ref{sec:perihelion}. In section~\ref{sec:annual}, we zoom out and consider the temporal evolution of water and its transport as well as the influence of GWs during the MY34 and MY28 dust storms. The results for the MY34 simulations are compared with observations by the Atmospheric Chemistry Suite (ACS) instrument onboard the ExoMars Trace Gas Orbiter spacecraft in section~\ref{sec:acs}.
	
	\section{Martian General Circulation Model and Design of Simulations}
	\label{sec:MGCM}
	
	The MPI--MGCM is based on a spectral dynamical core, which solves the three-dimensional nonlinear primitive equations of hydrodynamics on a globe. Physical parameterization suitable for the Martian atmosphere are described in detail in the papers of \citeA{hartogh2005description, hartogh2007middle, medvedev2007winter}. The vertical grid is represented by 67 hybrid $\eta$-levels extending from the surface to the pressure level $3.6\times 10^{-6}$ Pa ($\sim$160~km) in the thermosphere. All simulations were performed at the T21 horizontal resolution, which corresponds to $\approx 5.6^\circ\times 5.6^\circ$ horizontal grid. The subgrid-scale GW spectral parameterization was described in the work of \citeA{medvedev2011influence}, and its recent application to studying global effects of GWs is given in the work by \citeA{yiugit2018influence}. Other model parameters are the same as in the study of \citeA{medvedev2016comparison}.
	
	The water cycle of the model includes a semi-Lagrangian transport of water vapor and ice \cite{shaposhnikov2016water} and accounts for the microphysics of conversions between vapor and ice \cite{shaposhnikov2018modeling}. Condensation occurs on cloud condensation nuclei (CCN), whose sizes are represented by four characteristic bins. A bimodal log-normal dust size distribution is assumed in each spatial bin following the observations of \citeA{fedorova2014evidence}. The loss of vapor in the upper atmosphere is accounted for through photodissociation at the Lyman-$\alpha$ wavelength as well as by applying open boundary conditions at the model top. 
	
	Two predetermined dust scenarios for MY28 (2006--2007) \cite{medvedev2013general} and MY34 (2017--2019) \cite{montabone2020martian} were utilized in the simulations. The scenarios are based on the observed total dust optical depth in IR. Vertical distributions of the dust mixing ratio were calculated following the work by \citeA{conrath1975thermal} with modifications as described in the paper of \citeA[formulae (1)--(2)]{medvedev2013general}. Note a typo in (2) there: the expressions for $\tau\geq2$ and $\tau<2$ in (2) should be switched.
	
	\section{Influence of the MY34 Dust Storm}
	\label{sec:storm}
	
	We start with an overall assessment of the impact of the MY34 dust storm on the circulation and hydrological cycle. We selected a 10-day period at the beginning of the storm ($L_s=194^\circ - 200^\circ$), when changes in the atmospheric fields were most rapid. For comparison, we use the same dates from the simulation for MY28, when the storm has not started yet (see section~\ref{sec:annual}) and, thus, can be considered as a characteristic for dustless conditions. Note that the MY34 and MY28 runs include parameterized effects of subgrid-scale GWs as a standard model setup. Figure~\ref{fig:storm} presents the temperature and water vapor abundance for MY34 (contours) and the associated differences between the simulations for MY34 and MY28 (shaded). Figures~\ref{fig:storm}a,c,e demonstrate the atmospheric warming between 10~and 80~km by up to 30~K caused by the increased absorption of the solar radiation by airborne aerosol, especially in low latitudes. Dust particles block the radiation from reaching the surface, thus cooling down the lowermost layers. Changes in the meridional circulation lead to an adiabatic cooling in the upper atmosphere and heating in high latitudes (the so-called middle atmosphere polar warmings), as was discussed in the paper of \citeA[Sect. 5]{medvedev2013general}. 
	The dust-induced changes in the water vapor distribution are plotted in Figure~\ref{fig:storm}b, while the net budget of water is presented in Figures~\ref{fig:storm}d,f. It shows that the amount of vapor in the lower atmosphere significantly drops in response to stronger condensation at colder air temperature as well as due to a weaker sublimation from the surface. Contrary to that, higher temperature in the middle atmosphere leads to an enhanced saturation and increase of vapor at 40--80 km, especially in low latitudes. Finally, the comparison shows that the dust storm produces strong water pumping into the upper atmosphere, similar to that during the solstitial dust storm of MY28 \cite{shaposhnikov2019seasonal}. However, the main transmission channel in this case is over the equator, unlike over the polar region at perihelion (see the paper of \citeA{neary2020explanation} and section~\ref{sec:perihelion} for more discussion). Note that water is not lifted up above 40 km under the dustless scenario. Once vapor reaches the middle and upper atmosphere during the equinoctial MY34 storm, it is rapidly transported over all latitudes by the two-cell meridional circulation.
	
	A brief comparison of the simulated temperature with available measurements by MCS--MRO is given in Figures~\ref{fig:storm}c,e. The contours present the data from NASA's Planetary Data System \cite{mcs2021}. \change{It is seen}{We find} that the simulated temperatures are 5 to 10~K smaller than the retrieved ones, which may result in an underestimation by the model of the strength of the meridional cell in the middle atmosphere in the northern hemisphere and of the intensity of downwelling near the north pole, in particular.
	
	\section{Roles of Advection and Molecular Diffusion in the Upward Water Transport} 
	\label{sec:adv-diff}
	
	In order to further investigate pathways by which water is transported vertically we assess the relative roles of vertical advection and molecular diffusion. For that, we selected a day before the onset of the MY34 dust storm ($L_s\approx 180^\circ$) and at its midst ($L_s\approx 200^\circ$), and calculated the corresponding terms from the conservation equation for vapor: $wdq/dz$ and $\rho^{-1}d/dz(\mu dq/dz)$, where $w$ is the vertical velocity, $q$ is the water vapor mixing ratio, $\rho$ is the air density and $\mu$ is molecular dynamic viscosity. They characterize the rate of change $\partial q/\partial t$ by both processes, and are plotted in Figure~\ref{fig:diffusion} with color shades. Division of these quantities by $q$ yields inverse characteristic time scales, which are also shown with contours. It is seen that advection dominates the transport in the lower and middle atmosphere, while the role of molecular diffusion is appreciable above the mesopause and increases with height. Next, both the advective and diffusive vertical transport enhance by approximately an order of magnitude during the storm, as depicted by the scales of color bars in the upper and lower rows. With and without dust storms, advective time scales are much shorter than the diffusive ones (weeks and months vs years) below $\sim$100 km. Between $\sim$100 and 120 km, the time scales are of the same order, thus signifying the importance of both transport mechanisms. Above 120 km, molecular diffusion takes over vertical advection. The most important inference in the context of this study is that water is transported across the “bottleneck” (60-80 km) to the mesopause region ($\sim$110 km) by advection, while molecular diffusion plays virtually no role. Note that this conclusion does not concern horizontal transport, whose characteristic time scales between 60 and 120 km are much shorter (weeks and days), and which is solely responsible for latitudinal re-distribution of vapor. 
	
	\section{Effects of Gravity Waves on Water Transport During the MY34 Dust Storm}
	\label{sec:perihelion}
	
	Figuring out the primary role of the global circulation (advection), we next consider how GWs affect it and water transport during the equinoctial MY34 dust storm. This is done by comparing two simulations with the enabled and disabled GW parameterization (``GW on~and off") for the same as in section~\ref{sec:storm} period, i.e., the beginning of the storm ($L_s=194^\circ - 200^\circ$). The meridional circulation in these two cases is visualized with stream functions in the latitude-altitude plane in Figures~\ref{fig:perihelion}a (``GW on") and \ref{fig:perihelion}b (``GW off"). The arrows show the direction of the zonal mean Eulerian transport of air and water vapor parcels, and the red and blue contours indicate the clockwise and anti-clockwise circulation cells, respectively. It is seen that close to the equinox, the two cells are almost symmetric (with respect to the equator) in the lower and middle atmosphere. The major differences caused by GWs occur near and above the mesopause ($\sim$100-120 km, depending on the latitude). Momentum forcing by breaking/dissipating GWs a) intensifies the upper portions of both cells (the meridional velocity is proportional to the vertical distance between the streamlines), and b) reverses the meridional circulation inducing the one-cell south-to-north transport. The consequences for the water flux are shown in the same panels. Red and blue shades present the differences between the scenarios with and without GWs for the upward (Figure~\ref{fig:perihelion}a) and downward (Figure~\ref{fig:perihelion}b) fluxes.
	It is seen that the enhancement of the two-cell circulation increases the upward flux of water in low- to middle latitudes between $\sim$50 and 100 km (reddish shades in panel a) and the downward fluxes in the polar regions of both hemispheres at approximately the same heights (red shades in panel b). In the thermosphere above 110-120 km, the GW-induced pole-to-pole cell reduces the vertical water flux in low latitudes (blue shades in panel a), but speeds up the meridional transport.
	
	The most discussed, and thus recognizable, effect of GWs is acceleration/deceleration of the mean zonal wind, the latitude-altitude cross-sections of which are plotted in Figures~\ref{fig:perihelion}c and \ref{fig:perihelion}d for the simulations with and without the subgrid-scale GWs, correspondingly. The equinoctial circulation is represented by two eastward (prograde) midlatitude jets, with the one in the northern hemisphere being stronger due to the seasonal transition to the northern hemisphere fall. Both jets extend much higher, if the GW momentum forcing, or GW drag, $a_x$ (shown with contours in Figure~\ref{fig:perihelion}c) is not taken into account. Overall, the mean GW drag is directed against the mean wind. It affects mainly the upper parts of the jets, reduces their strength and even reverses the direction of the flow in the mesosphere and lower thermosphere. The other dynamically important effect of the GW drag is maintaining the meridional circulation. Outside the equatorial region, the mean meridional transport velocity is $\bar{v}^* \approx -(2\Omega\sin \varphi)^{-1} a_x$, where $\Omega$ is the rotational rate of the planet and $\varphi$ is the latitude \cite<e.g., see the discussion around Eqn. (5) in>[]{medvedev2007winter}. The additional dynamical forcing by the small-scale GWs of hundreds m~s$^{-1}$~sol$^{-1}$ accelerates the meridional flow in the northern hemisphere and reverses it in the southern hemisphere above the mesopause, according to the formula and as depicted by streamlines in Figure~\ref{fig:perihelion}a. This has important consequences for water entering this region: the meridional transport rapidly re-distributes it over the globe.
	
	The latter is illustrated in Figure~\ref{fig:perihelion}e, which shows the increased (by 40 to 60 ppm) water vapor mixing ratio in the thermosphere and its spread across latitudes, if GW forcing is taken into account. Due to continuity, the enhancement of the horizontal flow is accompanied by an intensification of the downward motions over the poles, especially in the northern hemisphere. The associated adiabatic heating affects temperature and produces the so-called polar warmings, with the northern one being stronger and more extensive (Figure~\ref{fig:perihelion}f). Warmer air and the influx of water support more vapor in polar regions, again more in the northern hemisphere. In turn, the intensified lower atmospheric branches of the circulation transport more water equatorward and facilitate its vertical pump in low latitudes. The colder air at low and middle latitudes above $\sim$70 km is the result of the enhanced vertical motions and the associated adiabatic cooling. This explains more water ice (shown with gray contours in Figures~\ref{fig:perihelion}a,b) in the simulation with GWs. Finally, the drop of the water vapor abundance just below is also the result	of its intense upward transport.
	
	Considering only the beginning of~the MY34 dust storm may create an impression that the impact of~GWs on water vapor is characterized exclusively by an increase in water concentration in the upper atmosphere. In the next section we explore the seasonal evolution of the vapor flux to show the complexity of GW effects on the distribution of water.
	
	\section{Influence of Gravity Waves on the Seasonal Behavior of Water During Dust Storms}
	\label{sec:annual}
	
	We compare temporal variations of water vapor during two major dust storms of MY34 and MY28. The former took place close to the equinox, the latter has developed near the solstice, hence, we call them the equinoctial and solstitial storms, respectively. The corresponding observed total dust optical depth in IR is shown in Figures~\ref{fig:annual}a and \ref{fig:annual}b with contours. Since our focus is on the penetration of water into the upper atmosphere taking into account GW effects, we plotted with shades the vertical flux of water vapor at~80~km in the same panels. Water usually does not rise that high throughout the year, therefore, this altitude can be regarded as a bottleneck. Positive (red) and negative (blue) values indicate the upward and downward fluxes. It is seen that, in both cases, a rapid increase of vertical fluxes of H$_2$O coincides with the onsets of the dust storms, with maxima collocated approximately with those of airborne dust at low-latitudes. The fluxes decline when the dust events settle down. The differences in latitudinal distributions between the two cases reflect the types of the meridional circulation. During the equinoctial MY34 storm, the upward branch of the circulation cell is in low latitudes, and most of water enters the upper atmosphere near the equator. Rapidly increased downward water fluxes in middle to high latitudes in both hemispheres (dark blue shades) are due to the intensified returning branches of the two-cell circulation. Both upward and downward fluxes gradually diminish, as the storm declines. A similar but weaker pattern repeats later this year (around $L_s=330^\circ$) during the minor equinoctial storm.
	
	The development of the global MY28 storm was less rapid than the one in MY34, and consisted of a series of minor storms, during which the meridional circulation itself seasonally transited from the equinoctial two-cell to the one-cell pole-to-pole type. It was initiated at $L_s=261.7^\circ$, probably driven by a Chryse storm track flushing storm \cite{wang2015origin}. The transition to solstitial circulation was accelerated by a regional dust storm occurred around $L_s=220^\circ$ followed by another one at $L_s=235^\circ$. Between $L_s=240-260^\circ$, Mars was free of regional dust storm activity. The peak of the planet-encircling storm occurred around the solstice ($L_s\approx 270^\circ$). The MY28 and 34 storms do contrast somewhat, with lifting centered near 50$^\circ$S during MY28 compared to MY34, where lifting centers were largely equatorial \cite{heavens2019dusty, heavens2019observational}. 
	
	Again, the distribution of vertical water fluxes reflects that of the meridional circulation with the maximum of the updraft in high latitudes of the southern (summer) hemisphere. The amount of water is usually much lower in the winter hemisphere. However, Figure~\ref{fig:annual}b shows very large downward water fluxes there. They are the result of the combination of a) the dust storm-enhanced downward motions and b) the increased concentration of vapor transported from the southern hemisphere. Thus, the main channels, through which water is delivered from the lower to the upper atmosphere coincides with the location of the upward branch of the meridional circulation: in low latitudes during equinoxes (MY34) and in high latitudes of the summer hemisphere (MY28). A certain amount of water is returned back to the middle and lower atmosphere by the global circulation cells, thus decreasing the total amount of vapor in the upper atmosphere.      
	
	Having considered the general patterns of water transport during dust storms, we now turn attention to the influence of GWs in this process. For that, we examine the globally averaged differences in water vapor content from simulations with and without GWs, which are shown with shades in Figures~\ref{fig:annual}c and \ref{fig:annual}d for MY34 and MY28, correspondingly. The net budget of water abundances in these simulations is also shown in Figure~\ref{fig:annual}. It is seen that the account of GWs leads to up to 50 ppmv (in the global-mean sense) increase of water above 100 km during the initial phase (up to $L_s\approx 220^\circ$) of the MY34 dust storm. This represents up to 15\% change of the total amount (Figure~\ref{fig:annual}e,g). After that and up to the end of the storm, the decrease of a similar magnitude (15-20\% ) is seen above $\sim$120 km. The analysis shows that such behavior is caused by water flux differences in low and middle latitudes, in the first hand. During the initial phase of the storm, they are up to 15 ppmv~m~s$^{-1}$ larger, if the GW forcing is included. We also note the increase of global water content between 40 and 70 km. Comparison with Figure~\ref{fig:perihelion}e shows that it occurs due to accumulation of vapor in the high-latitude regions. Similarly, the globally averaged temperature differences (contours in Figure~\ref{fig:annual}c) are the result of the GW-induced polar warming at these altitudes (see also Figure~\ref{fig:perihelion}f). Higher temperatures facilitate the vapor build-up by preventing condensation. Contrary to that, the GW-induced cooling above 80 km does not affect vapor directly due to the lack of nuclei and condensation.
	
	Later in the season of MY34 (between $L_s\approx 220^\circ$ and 270$^\circ$), the meridional circulation cell becomes flatter, and the influx of water from below in low latitudes stops. This occurs first in the upper portion of the model domain (above 120 km) and then spreads downward. The GW-enhanced returning branch of the circulation cell in high latitudes of the northern hemisphere facilitates the removal of water from the atmosphere above 120 km, but this process affects the total amount to a lesser degree. 
	
	A sudden drop of water (compared to the run without GWs) above 100 km during the gradual development of the MY28 dust storm precedes the rapid rise afterwards (Figures~\ref{fig:annual}f). This difference represents up to 25\% change, which can be clearly seen in Figure~\ref{fig:annual}f,h. The reason for this GW-induced drop can be found from analyzing the vertical fluxes in Figure~\ref{fig:annual}b. Before approximately $L_s=260^\circ$, the meridional circulation transitioned from the equinoctial type to the global one-cell solstitial one. GW forcing enhanced the downward part of the cell, thereby limiting the flow of water upward. With the onset of the global storm around $L_s=270^\circ$, the circulation reversed, and GWs enhanced the vertical water flux.
	
	In summary, dust storm-induced changes in the meridional circulation are primarily responsible for the enhanced upward water transport. GW forcing plays a secondary but important dynamical role in modulating the circulation and transport, thus affecting timing and leading up to 10-25\% changes in terms of the global water content in the upper atmosphere.
	
	\section{Comparison with ACS Observations}
	\label{sec:acs}
	
	Simulations of water with the MPI--MGCM for MY28 have been previously validated \cite{shaposhnikov2019seasonal} by comparing with the data obtained from the Mars Climate Sounder instrument onboard Mars Reconnaissance Orbiter (MCS--MRO) \cite{heavens2018hydrogen}. In this section, we compare the simulations for MY34 with observations performed by Atmospheric Chemistry Suite (ACS) onboard ExoMars Trace Gas Orbiter. ACS is~an~assembly of~three infrared spectrometers, two of which - the near-infrared (NIR, 0.7 to~1.7 $\mu$m) and middle-infrared (MIR, around 2.6 - 2.7 $\mu$m) channels were used for retrieving temperature and water vapor \cite{fedorova2020stormy, Belyaev_etal21}. Figures~\ref{fig:acs}a and \ref{fig:acs}c present the latitude-altitude distributions of temperature and water vapor, correspondingly, compiled from 638 measurements between $L_s=185^\circ$ and $267^\circ$. This interval covers the period of the MY34 major dust storm. The MGCM data were selected in a similar manner to match the spatiotemporal coverage of the ACS observations, and the averages are plotted in Figures~\ref{fig:acs}b and \ref{fig:acs}d. The individual profiles vary greatly over time in the simulations, therefore we present also the plots composed of minimums (Figure~\ref{fig:acs}e) and maximums (Figure~\ref{fig:acs}f) from the selected bins.
	
	It is seen that the model reproduces the observed temperature, generally, well. This includes both values and latitude-altitude distribution. Nevertheless, we note systematic temperature differences in the equatorial atmosphere at $\sim$75~km. The local minimum in the data at $\sim$35~km at 30-45$^\circ$S and differences in the temperature structure near the north polar hood can be caused by the small amount of ACS profiles on the observational borders. Diagonal banding in the data and the model at mid-high altitudes generally look similar.
	
	The model also reproduces the observed large-scale latitudinal and vertical structure of the water vapor, although the simulated abundances are larger than the measurements. The agreement is much better for the lower water vapor estimate (Figure~\ref{fig:acs}e). Given the sensitivity of the hydrological model to microphysics parameters, this can be related, in particular, to the prescribed dust scenario. Both observations and simulations show less than 10 ppmv of water between 20 and 30 km at polar regions. The maximum of abundances are located around 50 km near the equator. Above that height (between 60 and 70 km), water vapor decreases in this region, but not at middle to high latitudes, both in the observations and simulations. There are differences in the vertical structure of temperature near the south pole, with distinct diagonal pattern and a high altitude patch in the model that are present to a lesser extent in the ACS data. Vertical bands of large water vapor abundances in the model also are not present in the data, particularly at $\sim$60-80$^\circ$N. Note, that these differences are located mostly on the observational borders. The model predicts more vapor near the top of the domain. High volume mixing ratios in the thin atmosphere still correspond to low concentration of molecules. This is why the instrumental errors of ACS are large there. The other potential source of differences can be the lack of a detailed water photochemistry in the model, which is very complex in the upper atmosphere \cite{Stone_etal20}. Inclusion of it in the MGCM would \change{almost certainly}{likely} bring the simulated water closer to the observations.
	
	\section{Summary and Conclusions}
	\label{sec:conclusions}
	
	We have examined the hydrological cycle on Mars during two major dust storms of MY28 and 34 using the MPI Martian general circulation model and predetermined dust scenarios. A particular focus was on the impact of small-scale gravity waves (GWs), which were accounted for in the model with the nonlinear spectral parameterization \cite{Yigit_etal08, medvedev2011influence}. The simulations further elucidated the mechanism of water transport to the upper atmosphere by the meridional circulation - the so-called global water ``pump" \cite{shaposhnikov2019seasonal}. Dust storms and GW forcing affect the circulation and temperature and, thus, control the amount of water vapor in the thermosphere, its subsequent photodissociation and escape of hydrogen to space. Although changes in temperature and circulation are intertwined, their influence on water is different in the lower and upper layers. In the lower atmosphere, warmer air can sustain more vapor and facilitates its sublimation from the reservoirs on the surface. In the upper atmosphere, the lack of nuclei prevents condensation, and water can only be redistributed. The main findings of this work concerning the impact of global dust storms and GWs on the water cycle are summarized below.
	
	\begin{enumerate}
		\item Water is transported above 80-100 km by the global circulation (advection), with molecular diffusion playing virtually no role. The impact of the latter grows with height, and can no longer be neglected above $\sim$120 km.
		\item The main channels, through which water enters the upper atmosphere, are collocated with the regions of air updraft by the meridional circulation: in low latitudes during the northern autumn equinox and in high southern latitudes during perihelion.
		\item Dust storm-induced global circulation is the primary mechanism that enables water penetration to the upper atmosphere. 
		\item GW forcing plays an important role in shaping the distribution of water by modulating the timing and intensity of the transport.
		\item Accounting for small-scale GWs contributes to changes in globally averaged high-altitude water abundance of up to 10-25\%. 
		\item During the equinoctial dust storm (MY34), the GW forcing leads to a globally averaged increase (by up to 50 ppmv) of water above 80 km in the beginning of the storm, more rapid drop at the later stage, and accumulation of vapor in polar regions of both hemispheres in the lower and middle atmosphere.     
		\item During the solstitial major dust event (MY28), the impact of GWs intensifies the effect of the storm by increasing the high-altitude water amount by more than 50 ppmv (in the global mean sense). GWs can  produce a similarly large drop of water during the earlier stage of the storm.
	\end{enumerate}
	
	The presented MGCM simulations and comparison with ACS solidify the concept of the water ``pump" mechanism of transporting vapor directly to the upper atmosphere. Model simulations also provide finer details and predictions, which can be later verified with observations, once more data become available.
	Overall, our results highlight the wide-reaching implications of the GW-induced large-scale circulation and its potential role in modulating the distribution of water on Mars.
	
	\acknowledgments
	
	The data supporting the MPI--MGCM simulations can be found at \url{https://zenodo.org/record/5749676} \cite{dmitry_s_shaposhnikov_2021_5749676}.  The most recent model output can be accessed at \url{https://mars.mipt.ru}. The ACS MIR temperature and water data are available at \url{http://exomars.cosmos.ru/ACS_Results_stormy_water_vREzUd4pxG/} \cite{fedorova2020stormy} and \url{https://data.mendeley.com/datasets/995y7ymdgm/1} \cite{Belyaev_etal21}, correspondingly. The MCS--MRO data are available at \url{https://atmos.nmsu.edu/PDS/data} \cite{mcs2021}.
	
	The authors thank Denis Belyaev of IKI for assistance with the observational data and helpful discussions. The work was partially supported by the Russian Science Foundation grant 20-72-00110.
	
	
	\newpage
	
	
	\bibliography{maoam-jgr}

\begin{thebibliography}{}

\bibitem [\protect \citeauthoryear {%
Aoki%
\ \protect \BOthers {.}}{%
Aoki%
\ \protect \BOthers {.}}{%
{\protect \APACyear {2019}}%
}]{%
aoki2019water}
\APACinsertmetastar {%
aoki2019water}%
\begin{APACrefauthors}%
Aoki, S.%
, Vandaele, A\BPBI C.%
, Daerden, F.%
, Villanueva, G\BPBI L.%
, Liuzzi, G.%
, Thomas, I\BPBI R.%
\BDBL {}others%
\end{APACrefauthors}%
\unskip\
\newblock
\APACrefYearMonthDay{2019}{}{}.
\newblock
{\BBOQ}\APACrefatitle {Water vapor vertical profiles on {M}ars in dust storms
  observed by {TGO/NOMAD}} {Water vapor vertical profiles on {M}ars in dust
  storms observed by {TGO/NOMAD}}.{\BBCQ}
\newblock
\APACjournalVolNumPages{Journal of Geophysical Research:
  Planets}{124}{12}{3482--3497}.
\PrintBackRefs{\CurrentBib}

\bibitem [\protect \citeauthoryear {%
Belyaev%
\ \protect \BOthers {.}}{%
Belyaev%
\ \protect \BOthers {.}}{%
{\protect \APACyear {2021}}%
}]{%
Belyaev_etal21}
\APACinsertmetastar {%
Belyaev_etal21}%
\begin{APACrefauthors}%
Belyaev, D\BPBI A.%
, Fedorova, A\BPBI A.%
, Trokhimovskiy, A.%
, Alday, J.%
, Montmessin, F.%
, Korablev, O\BPBI I.%
\BDBL {}Shakun, A\BPBI V.%
\end{APACrefauthors}%
\unskip\
\newblock
\APACrefYearMonthDay{2021}{}{}.
\newblock
{\BBOQ}\APACrefatitle {Revealing a High Water Abundance in the Upper Mesosphere
  of {M}ars With {ACS} Onboard {TGO}} {Revealing a high water abundance in the
  upper mesosphere of {M}ars with {ACS} onboard {TGO}}.{\BBCQ}
\newblock
\APACjournalVolNumPages{Geophysical Research Letters}{48}{10}{e2021GL093411}.
\newblock
\begin{APACrefURL}
  \url{https://agupubs.onlinelibrary.wiley.com/doi/abs/10.1029/2021GL093411}
  \end{APACrefURL}
\newblock
\begin{APACrefDOI} \doi{10.1029/2021GL093411} \end{APACrefDOI}
\PrintBackRefs{\CurrentBib}

\bibitem [\protect \citeauthoryear {%
Bhattacharyya%
\ \protect \BOthers {.}}{%
Bhattacharyya%
\ \protect \BOthers {.}}{%
{\protect \APACyear {2017}}%
}]{%
bhattacharyya2017seasonal}
\APACinsertmetastar {%
bhattacharyya2017seasonal}%
\begin{APACrefauthors}%
Bhattacharyya, D.%
, Clarke, J.%
, Chaufray, J\BHBI Y.%
, Mayyasi, M.%
, Bertaux, J\BHBI L.%
, Chaffin, M.%
\BDBL {}Villanueva, G.%
\end{APACrefauthors}%
\unskip\
\newblock
\APACrefYearMonthDay{2017}{}{}.
\newblock
{\BBOQ}\APACrefatitle {Seasonal changes in hydrogen escape from {M}ars through
  analysis of {HST} observations of the {M}artian exosphere near perihelion}
  {Seasonal changes in hydrogen escape from {M}ars through analysis of {HST}
  observations of the {M}artian exosphere near perihelion}.{\BBCQ}
\newblock
\APACjournalVolNumPages{Journal of Geophysical Research: Space
  Physics}{122}{11}{}.
\PrintBackRefs{\CurrentBib}

\bibitem [\protect \citeauthoryear {%
Bhattacharyya%
, Clarke%
, Bertaux%
, Chaufray%
\BCBL {}\ \BBA {} Mayyasi%
}{%
Bhattacharyya%
\ \protect \BOthers {.}}{%
{\protect \APACyear {2015}}%
}]{%
bhattacharyya2015strong}
\APACinsertmetastar {%
bhattacharyya2015strong}%
\begin{APACrefauthors}%
Bhattacharyya, D.%
, Clarke, J\BPBI T.%
, Bertaux, J\BHBI L.%
, Chaufray, J\BHBI Y.%
\BCBL {}\ \BBA {} Mayyasi, M.%
\end{APACrefauthors}%
\unskip\
\newblock
\APACrefYearMonthDay{2015}{}{}.
\newblock
{\BBOQ}\APACrefatitle {A strong seasonal dependence in the {M}artian hydrogen
  exosphere} {A strong seasonal dependence in the {M}artian hydrogen
  exosphere}.{\BBCQ}
\newblock
\APACjournalVolNumPages{Geophysical Research Letters}{42}{20}{8678--8685}.
\PrintBackRefs{\CurrentBib}

\bibitem [\protect \citeauthoryear {%
Chaffin%
\ \protect \BOthers {.}}{%
Chaffin%
\ \protect \BOthers {.}}{%
{\protect \APACyear {2014}}%
}]{%
chaffin2014unexpected}
\APACinsertmetastar {%
chaffin2014unexpected}%
\begin{APACrefauthors}%
Chaffin, M\BPBI S.%
, Chaufray, J\BHBI Y.%
, Stewart, I.%
, Montmessin, F.%
, Schneider, N\BPBI M.%
\BCBL {}\ \BBA {} Bertaux, J\BHBI L.%
\end{APACrefauthors}%
\unskip\
\newblock
\APACrefYearMonthDay{2014}{}{}.
\newblock
{\BBOQ}\APACrefatitle {Unexpected variability of Martian hydrogen escape}
  {Unexpected variability of martian hydrogen escape}.{\BBCQ}
\newblock
\APACjournalVolNumPages{Geophysical Research Letters}{41}{2}{314--320}.
\PrintBackRefs{\CurrentBib}

\bibitem [\protect \citeauthoryear {%
Chaffin%
, Deighan%
, Schneider%
\BCBL {}\ \BBA {} Stewart%
}{%
Chaffin%
\ \protect \BOthers {.}}{%
{\protect \APACyear {2017}}%
}]{%
chaffin2017elevated}
\APACinsertmetastar {%
chaffin2017elevated}%
\begin{APACrefauthors}%
Chaffin, M\BPBI S.%
, Deighan, J.%
, Schneider, N.%
\BCBL {}\ \BBA {} Stewart, A.%
\end{APACrefauthors}%
\unskip\
\newblock
\APACrefYearMonthDay{2017}{}{}.
\newblock
{\BBOQ}\APACrefatitle {Elevated atmospheric escape of atomic hydrogen from
  {M}ars induced by high-altitude water} {Elevated atmospheric escape of atomic
  hydrogen from {M}ars induced by high-altitude water}.{\BBCQ}
\newblock
\APACjournalVolNumPages{Nature Geoscience}{10}{3}{174}.
\PrintBackRefs{\CurrentBib}

\bibitem [\protect \citeauthoryear {%
Chaufray%
\ \protect \BOthers {.}}{%
Chaufray%
\ \protect \BOthers {.}}{%
{\protect \APACyear {2021}}%
}]{%
chaufray_etal21}
\APACinsertmetastar {%
chaufray_etal21}%
\begin{APACrefauthors}%
Chaufray, J\BHBI Y.%
, Gonzalez-Galindo, F.%
, Lopez-Valverde, M.%
, Forget, F.%
, Quémerais, E.%
, Bertaux, J\BHBI L.%
\BDBL {}Yelle, R.%
\end{APACrefauthors}%
\unskip\
\newblock
\APACrefYearMonthDay{2021}{}{}.
\newblock
{\BBOQ}\APACrefatitle {Study of the hydrogen escape rate at Mars during martian
  years 28 and 29 from comparisons between {SPICAM/M}ars express observations
  and {GCM-LMD} simulations} {Study of the hydrogen escape rate at mars during
  martian years 28 and 29 from comparisons between {SPICAM/M}ars express
  observations and {GCM-LMD} simulations}.{\BBCQ}
\newblock
\APACjournalVolNumPages{Icarus}{353}{}{113498}.
\newblock
\begin{APACrefURL}
  \url{https://www.sciencedirect.com/science/article/pii/S0019103518306985}
  \end{APACrefURL}
\newblock
\begin{APACrefDOI} \doi{https://doi.org/10.1016/j.icarus.2019.113498}
  \end{APACrefDOI}
\PrintBackRefs{\CurrentBib}

\bibitem [\protect \citeauthoryear {%
Clarke%
\ \protect \BOthers {.}}{%
Clarke%
\ \protect \BOthers {.}}{%
{\protect \APACyear {2014}}%
}]{%
clarke2014rapid}
\APACinsertmetastar {%
clarke2014rapid}%
\begin{APACrefauthors}%
Clarke, J\BPBI T.%
, Bertaux, J\BHBI L.%
, Chaufray, J\BHBI Y.%
, Gladstone, G\BPBI R.%
, Qu{\'e}merais, E.%
, Wilson, J.%
\BCBL {}\ \BBA {} Bhattacharyya, D.%
\end{APACrefauthors}%
\unskip\
\newblock
\APACrefYearMonthDay{2014}{}{}.
\newblock
{\BBOQ}\APACrefatitle {A rapid decrease of the hydrogen corona of {M}ars} {A
  rapid decrease of the hydrogen corona of {M}ars}.{\BBCQ}
\newblock
\APACjournalVolNumPages{Geophysical Research Letters}{41}{22}{8013--8020}.
\PrintBackRefs{\CurrentBib}

\bibitem [\protect \citeauthoryear {%
Conrath%
}{%
Conrath%
}{%
{\protect \APACyear {1975}}%
}]{%
conrath1975thermal}
\APACinsertmetastar {%
conrath1975thermal}%
\begin{APACrefauthors}%
Conrath, B\BPBI J.%
\end{APACrefauthors}%
\unskip\
\newblock
\APACrefYearMonthDay{1975}{}{}.
\newblock
{\BBOQ}\APACrefatitle {Thermal structure of the {M}artian atmosphere during the
  dissipation of the dust storm of 1971} {Thermal structure of the {M}artian
  atmosphere during the dissipation of the dust storm of 1971}.{\BBCQ}
\newblock
\APACjournalVolNumPages{Icarus}{24}{1}{36--46}.
\PrintBackRefs{\CurrentBib}

\bibitem [\protect \citeauthoryear {%
Fedorova%
\ \protect \BOthers {.}}{%
Fedorova%
\ \protect \BOthers {.}}{%
{\protect \APACyear {2018}}%
}]{%
fedorova2018water}
\APACinsertmetastar {%
fedorova2018water}%
\begin{APACrefauthors}%
Fedorova, A\BPBI A.%
, Bertaux, J\BHBI L.%
, Betsis, D.%
, Montmessin, F.%
, Korablev, O.%
, Maltagliati, L.%
\BCBL {}\ \BBA {} Clarke, J.%
\end{APACrefauthors}%
\unskip\
\newblock
\APACrefYearMonthDay{2018}{}{}.
\newblock
{\BBOQ}\APACrefatitle {Water vapor in the middle atmosphere of {M}ars during
  the 2007 global dust storm} {Water vapor in the middle atmosphere of {M}ars
  during the 2007 global dust storm}.{\BBCQ}
\newblock
\APACjournalVolNumPages{Icarus}{300}{}{440--457}.
\PrintBackRefs{\CurrentBib}

\bibitem [\protect \citeauthoryear {%
Fedorova%
\ \protect \BOthers {.}}{%
Fedorova%
\ \protect \BOthers {.}}{%
{\protect \APACyear {2021}}%
}]{%
Fedorova_etal21}
\APACinsertmetastar {%
Fedorova_etal21}%
\begin{APACrefauthors}%
Fedorova, A\BPBI A.%
, Montmessin, F.%
, Korablev, O.%
, Lefèvre, F.%
, Trokhimovskiy, A.%
\BCBL {}\ \BBA {} Bertaux, J\BHBI L.%
\end{APACrefauthors}%
\unskip\
\newblock
\APACrefYearMonthDay{2021}{}{}.
\newblock
{\BBOQ}\APACrefatitle {Multi-Annual Monitoring of the Water Vapor Vertical
  Distribution on {M}ars by {SPICAM} on {M}ars {E}xpress} {Multi-annual
  monitoring of the water vapor vertical distribution on {M}ars by {SPICAM} on
  {M}ars {E}xpress}.{\BBCQ}
\newblock
\APACjournalVolNumPages{Journal of Geophysical Research:
  Planets}{126}{}{e2020JE006616}.
\newblock
\begin{APACrefDOI} \doi{10.1029/2020JE006616} \end{APACrefDOI}
\PrintBackRefs{\CurrentBib}

\bibitem [\protect \citeauthoryear {%
Fedorova%
\ \protect \BOthers {.}}{%
Fedorova%
\ \protect \BOthers {.}}{%
{\protect \APACyear {2020}}%
}]{%
fedorova2020stormy}
\APACinsertmetastar {%
fedorova2020stormy}%
\begin{APACrefauthors}%
Fedorova, A\BPBI A.%
, Montmessin, F.%
, Korablev, O.%
, Luginin, M.%
, Trokhimovskiy, A.%
, Belyaev, D\BPBI A.%
\BDBL {}others%
\end{APACrefauthors}%
\unskip\
\newblock
\APACrefYearMonthDay{2020}{}{}.
\newblock
{\BBOQ}\APACrefatitle {Stormy water on {M}ars: The distribution and saturation
  of atmospheric water during the dusty season} {Stormy water on {M}ars: The
  distribution and saturation of atmospheric water during the dusty
  season}.{\BBCQ}
\newblock
\APACjournalVolNumPages{Science}{367}{6475}{297--300}.
\PrintBackRefs{\CurrentBib}

\bibitem [\protect \citeauthoryear {%
Fedorova%
\ \protect \BOthers {.}}{%
Fedorova%
\ \protect \BOthers {.}}{%
{\protect \APACyear {2014}}%
}]{%
fedorova2014evidence}
\APACinsertmetastar {%
fedorova2014evidence}%
\begin{APACrefauthors}%
Fedorova, A\BPBI A.%
, Montmessin, F.%
, Rodin, A\BPBI V.%
, Korablev, O\BPBI I.%
, M{\"{a}}{\"{a}}tt{\"{a}}nen, A.%
, Maltagliati, L.%
\BCBL {}\ \BBA {} Bertaux, J\BPBI L.%
\end{APACrefauthors}%
\unskip\
\newblock
\APACrefYearMonthDay{2014}{}{}.
\newblock
{\BBOQ}\APACrefatitle {{Evidence for a bimodal size distribution for the
  suspended aerosol particles on Mars}} {{Evidence for a bimodal size
  distribution for the suspended aerosol particles on Mars}}.{\BBCQ}
\newblock
\APACjournalVolNumPages{Icarus}{231}{}{239--260}.
\newblock
\begin{APACrefDOI} \doi{10.1016/j.icarus.2013.12.015} \end{APACrefDOI}
\PrintBackRefs{\CurrentBib}

\bibitem [\protect \citeauthoryear {%
Halekas%
}{%
Halekas%
}{%
{\protect \APACyear {2017}}%
}]{%
halekas2017seasonal}
\APACinsertmetastar {%
halekas2017seasonal}%
\begin{APACrefauthors}%
Halekas, J.%
\end{APACrefauthors}%
\unskip\
\newblock
\APACrefYearMonthDay{2017}{}{}.
\newblock
{\BBOQ}\APACrefatitle {Seasonal variability of the hydrogen exosphere of
  {M}ars} {Seasonal variability of the hydrogen exosphere of {M}ars}.{\BBCQ}
\newblock
\APACjournalVolNumPages{Journal of Geophysical Research:
  Planets}{122}{5}{901--911}.
\PrintBackRefs{\CurrentBib}

\bibitem [\protect \citeauthoryear {%
Hartogh%
, Medvedev%
\BCBL {}\ \BBA {} Jarchow%
}{%
Hartogh%
\ \protect \BOthers {.}}{%
{\protect \APACyear {2007}}%
}]{%
hartogh2007middle}
\APACinsertmetastar {%
hartogh2007middle}%
\begin{APACrefauthors}%
Hartogh, P.%
, Medvedev, A\BPBI S.%
\BCBL {}\ \BBA {} Jarchow, C.%
\end{APACrefauthors}%
\unskip\
\newblock
\APACrefYearMonthDay{2007}{}{}.
\newblock
{\BBOQ}\APACrefatitle {{Middle atmosphere polar warmings on Mars: Simulations
  and study on the validation with sub-millimeter observations}} {{Middle
  atmosphere polar warmings on Mars: Simulations and study on the validation
  with sub-millimeter observations}}.{\BBCQ}
\newblock
\APACjournalVolNumPages{Planetary and Space Science}{55}{9}{1103--1112}.
\newblock
\begin{APACrefDOI} \doi{10.1016/j.pss.2006.11.018} \end{APACrefDOI}
\PrintBackRefs{\CurrentBib}

\bibitem [\protect \citeauthoryear {%
Hartogh%
\ \protect \BOthers {.}}{%
Hartogh%
\ \protect \BOthers {.}}{%
{\protect \APACyear {2005}}%
}]{%
hartogh2005description}
\APACinsertmetastar {%
hartogh2005description}%
\begin{APACrefauthors}%
Hartogh, P.%
, Medvedev, A\BPBI S.%
, Kuroda, T.%
, Saito, R.%
, Villanueva, G.%
, Feofilov, A\BPBI G.%
\BDBL {}Berger, U.%
\end{APACrefauthors}%
\unskip\
\newblock
\APACrefYearMonthDay{2005}{}{}.
\newblock
{\BBOQ}\APACrefatitle {{Description and climatology of a new general
  circulation model of the Martian atmosphere}} {{Description and climatology
  of a new general circulation model of the Martian atmosphere}}.{\BBCQ}
\newblock
\APACjournalVolNumPages{Journal of Geophysical Research}{110}{E11}{E11008}.
\newblock
\begin{APACrefDOI} \doi{10.1029/2005JE002498} \end{APACrefDOI}
\PrintBackRefs{\CurrentBib}

\bibitem [\protect \citeauthoryear {%
Heavens%
, Kass%
, Kleinb{\"o}hl%
\BCBL {}\ \BBA {} Schofield%
}{%
Heavens%
\ \protect \BOthers {.}}{%
{\protect \APACyear {2020}}%
}]{%
heavens2020multiannual}
\APACinsertmetastar {%
heavens2020multiannual}%
\begin{APACrefauthors}%
Heavens, N\BPBI G.%
, Kass, D\BPBI M.%
, Kleinb{\"o}hl, A.%
\BCBL {}\ \BBA {} Schofield, J\BPBI T.%
\end{APACrefauthors}%
\unskip\
\newblock
\APACrefYearMonthDay{2020}{}{}.
\newblock
{\BBOQ}\APACrefatitle {A multiannual record of gravity wave activity in
  {M}ars’s lower atmosphere from on-planet observations by the {M}ars
  {C}limate {S}ounder} {A multiannual record of gravity wave activity in
  {M}ars’s lower atmosphere from on-planet observations by the {M}ars
  {C}limate {S}ounder}.{\BBCQ}
\newblock
\APACjournalVolNumPages{Icarus}{341}{}{113630}.
\PrintBackRefs{\CurrentBib}

\bibitem [\protect \citeauthoryear {%
Heavens%
, Kass%
\BCBL {}\ \BBA {} Shirley%
}{%
Heavens%
, Kass%
\BCBL {}\ \BBA {} Shirley%
}{%
{\protect \APACyear {2019}}%
}]{%
heavens2019dusty}
\APACinsertmetastar {%
heavens2019dusty}%
\begin{APACrefauthors}%
Heavens, N\BPBI G.%
, Kass, D\BPBI M.%
\BCBL {}\ \BBA {} Shirley, J\BPBI H.%
\end{APACrefauthors}%
\unskip\
\newblock
\APACrefYearMonthDay{2019}{}{}.
\newblock
{\BBOQ}\APACrefatitle {Dusty deep convection in the {M}ars year 34
  planet-encircling dust event} {Dusty deep convection in the {M}ars year 34
  planet-encircling dust event}.{\BBCQ}
\newblock
\APACjournalVolNumPages{Journal of Geophysical Research:
  Planets}{124}{11}{2863--2892}.
\PrintBackRefs{\CurrentBib}

\bibitem [\protect \citeauthoryear {%
Heavens%
, Kass%
, Shirley%
, Piqueux%
\BCBL {}\ \BBA {} Cantor%
}{%
Heavens%
, Kass%
, Shirley%
, Piqueux%
\BCBL {}\ \BBA {} Cantor%
}{%
{\protect \APACyear {2019}}%
}]{%
heavens2019observational}
\APACinsertmetastar {%
heavens2019observational}%
\begin{APACrefauthors}%
Heavens, N\BPBI G.%
, Kass, D\BPBI M.%
, Shirley, J\BPBI H.%
, Piqueux, S.%
\BCBL {}\ \BBA {} Cantor, B\BPBI A.%
\end{APACrefauthors}%
\unskip\
\newblock
\APACrefYearMonthDay{2019}{}{}.
\newblock
{\BBOQ}\APACrefatitle {An observational overview of dusty deep convection in
  {M}artian dust storms} {An observational overview of dusty deep convection in
  {M}artian dust storms}.{\BBCQ}
\newblock
\APACjournalVolNumPages{Journal of the atmospheric
  sciences}{76}{11}{3299--3326}.
\PrintBackRefs{\CurrentBib}

\bibitem [\protect \citeauthoryear {%
Heavens%
\ \protect \BOthers {.}}{%
Heavens%
\ \protect \BOthers {.}}{%
{\protect \APACyear {2018}}%
}]{%
heavens2018hydrogen}
\APACinsertmetastar {%
heavens2018hydrogen}%
\begin{APACrefauthors}%
Heavens, N\BPBI G.%
, Kleinb{\"o}hl, A.%
, Chaffin, M\BPBI S.%
, Halekas, J\BPBI S.%
, Kass, D\BPBI M.%
, Hayne, P\BPBI O.%
\BDBL {}Schofield, J\BPBI T.%
\end{APACrefauthors}%
\unskip\
\newblock
\APACrefYearMonthDay{2018}{}{}.
\newblock
{\BBOQ}\APACrefatitle {Hydrogen escape from {M}ars enhanced by deep convection
  in dust storms} {Hydrogen escape from {M}ars enhanced by deep convection in
  dust storms}.{\BBCQ}
\newblock
\APACjournalVolNumPages{Nature Astronomy}{2}{2}{126}.
\PrintBackRefs{\CurrentBib}

\bibitem [\protect \citeauthoryear {%
Hunten%
\ \BBA {} McElroy%
}{%
Hunten%
\ \BBA {} McElroy%
}{%
{\protect \APACyear {1970}}%
}]{%
hunten1970production}
\APACinsertmetastar {%
hunten1970production}%
\begin{APACrefauthors}%
Hunten, D.%
\BCBT {}\ \BBA {} McElroy, M.%
\end{APACrefauthors}%
\unskip\
\newblock
\APACrefYearMonthDay{1970}{}{}.
\newblock
{\BBOQ}\APACrefatitle {Production and escape of hydrogen on {M}ars} {Production
  and escape of hydrogen on {M}ars}.{\BBCQ}
\newblock
\APACjournalVolNumPages{Journal of Geophysical Research}{75}{31}{5989--6001}.
\PrintBackRefs{\CurrentBib}

\bibitem [\protect \citeauthoryear {%
Jesch%
, Medvedev%
, Castellini%
, Yi{\u{g}}it%
\BCBL {}\ \BBA {} Hartogh%
}{%
Jesch%
\ \protect \BOthers {.}}{%
{\protect \APACyear {2019}}%
}]{%
jesch2019density}
\APACinsertmetastar {%
jesch2019density}%
\begin{APACrefauthors}%
Jesch, D.%
, Medvedev, A\BPBI S.%
, Castellini, F.%
, Yi{\u{g}}it, E.%
\BCBL {}\ \BBA {} Hartogh, P.%
\end{APACrefauthors}%
\unskip\
\newblock
\APACrefYearMonthDay{2019}{}{}.
\newblock
{\BBOQ}\APACrefatitle {{Density fluctuations in the lower thermosphere of Mars
  retrieved from the ExoMars Trace Gas Orbiter (TGO) aerobraking}} {{Density
  fluctuations in the lower thermosphere of Mars retrieved from the ExoMars
  Trace Gas Orbiter (TGO) aerobraking}}.{\BBCQ}
\newblock
\APACjournalVolNumPages{Atmosphere}{10}{10}{620}.
\PrintBackRefs{\CurrentBib}

\bibitem [\protect \citeauthoryear {%
Leelavathi%
, Venkateswara~Rao%
\BCBL {}\ \BBA {} Rao%
}{%
Leelavathi%
\ \protect \BOthers {.}}{%
{\protect \APACyear {2020}}%
}]{%
Leelavathi_etal20}
\APACinsertmetastar {%
Leelavathi_etal20}%
\begin{APACrefauthors}%
Leelavathi, V.%
, Venkateswara~Rao, N.%
\BCBL {}\ \BBA {} Rao, S\BPBI V\BPBI B.%
\end{APACrefauthors}%
\unskip\
\newblock
\APACrefYearMonthDay{2020}{}{}.
\newblock
{\BBOQ}\APACrefatitle {Interannual Variability of Atmospheric Gravity Waves in
  the {M}artian Thermosphere: Effects of the 2018 Planet-Encircling Dust Event}
  {Interannual variability of atmospheric gravity waves in the {M}artian
  thermosphere: Effects of the 2018 planet-encircling dust event}.{\BBCQ}
\newblock
\APACjournalVolNumPages{Journal of Geophysical Research:
  Planets}{125}{12}{e2020JE006649}.
\newblock
\begin{APACrefDOI} \doi{10.1029/2020JE006649} \end{APACrefDOI}
\PrintBackRefs{\CurrentBib}

\bibitem [\protect \citeauthoryear {%
Li%
, Liu%
\BCBL {}\ \BBA {} Jin%
}{%
Li%
\ \protect \BOthers {.}}{%
{\protect \APACyear {2021}}%
}]{%
Li_etal21}
\APACinsertmetastar {%
Li_etal21}%
\begin{APACrefauthors}%
Li, Y.%
, Liu, J.%
\BCBL {}\ \BBA {} Jin, S.%
\end{APACrefauthors}%
\unskip\
\newblock
\APACrefYearMonthDay{2021}{}{}.
\newblock
{\BBOQ}\APACrefatitle {Horizontal Internal Gravity Waves in the {Mars} Upper
  Atmosphere From {MAVEN} {ACC} and {NGIMS} Measurements} {Horizontal internal
  gravity waves in the {Mars} upper atmosphere from {MAVEN} {ACC} and {NGIMS}
  measurements}.{\BBCQ}
\newblock
\APACjournalVolNumPages{Journal of Geophysical Research: Space
  Science}{126}{1}{e2020JA028378}.
\newblock
\begin{APACrefDOI} \doi{10.1029/2020JA028378} \end{APACrefDOI}
\PrintBackRefs{\CurrentBib}

\bibitem [\protect \citeauthoryear {%
Maltagliati%
\ \protect \BOthers {.}}{%
Maltagliati%
\ \protect \BOthers {.}}{%
{\protect \APACyear {2013}}%
}]{%
maltagliati2013annual}
\APACinsertmetastar {%
maltagliati2013annual}%
\begin{APACrefauthors}%
Maltagliati, L.%
, Montmessin, F.%
, Korablev, O.%
, Fedorova, A.%
, Forget, F.%
, M{\"a}{\"a}tt{\"a}nen, A.%
\BDBL {}Bertaux, J\BHBI L.%
\end{APACrefauthors}%
\unskip\
\newblock
\APACrefYearMonthDay{2013}{}{}.
\newblock
{\BBOQ}\APACrefatitle {Annual survey of water vapor vertical distribution and
  water--aerosol coupling in the martian atmosphere observed by {SPICAM/MEx}
  solar occultations} {Annual survey of water vapor vertical distribution and
  water--aerosol coupling in the martian atmosphere observed by {SPICAM/MEx}
  solar occultations}.{\BBCQ}
\newblock
\APACjournalVolNumPages{Icarus}{223}{2}{942--962}.
\PrintBackRefs{\CurrentBib}

\bibitem [\protect \citeauthoryear {%
Maltagliati%
\ \protect \BOthers {.}}{%
Maltagliati%
\ \protect \BOthers {.}}{%
{\protect \APACyear {2011}}%
}]{%
maltagliati2011annual}
\APACinsertmetastar {%
maltagliati2011annual}%
\begin{APACrefauthors}%
Maltagliati, L.%
, Titov, D\BPBI V.%
, Encrenaz, T.%
, Melchiorri, R.%
, Forget, F.%
, Keller, H\BPBI U.%
\BCBL {}\ \BBA {} Bibring, J\BPBI P.%
\end{APACrefauthors}%
\unskip\
\newblock
\APACrefYearMonthDay{2011}{}{}.
\newblock
{\BBOQ}\APACrefatitle {{Annual survey of water vapor behavior from the OMEGA
  mapping spectrometer onboard Mars Express}} {{Annual survey of water vapor
  behavior from the OMEGA mapping spectrometer onboard Mars Express}}.{\BBCQ}
\newblock
\APACjournalVolNumPages{Icarus}{213}{2}{480--495}.
\newblock
\begin{APACrefDOI} \doi{10.1016/j.icarus.2011.03.030} \end{APACrefDOI}
\PrintBackRefs{\CurrentBib}

\bibitem [\protect \citeauthoryear {%
McElroy%
}{%
McElroy%
}{%
{\protect \APACyear {1972}}%
}]{%
mcelroy1972mars}
\APACinsertmetastar {%
mcelroy1972mars}%
\begin{APACrefauthors}%
McElroy, M\BPBI B.%
\end{APACrefauthors}%
\unskip\
\newblock
\APACrefYearMonthDay{1972}{}{}.
\newblock
{\BBOQ}\APACrefatitle {Mars: An evolving atmosphere} {Mars: An evolving
  atmosphere}.{\BBCQ}
\newblock
\APACjournalVolNumPages{Science}{175}{4020}{443--445}.
\PrintBackRefs{\CurrentBib}

\bibitem [\protect \citeauthoryear {%
\APACcitebtitle {MCS {D}ata {A}rchive}}{%
\APACcitebtitle {MCS {D}ata {A}rchive}}{%
{\protect \APACyear {2021}}%
}]{%
mcs2021}
\APACinsertmetastar {%
mcs2021}%
\APACrefbtitle {MCS {D}ata {A}rchive.} {Mcs {D}ata {A}rchive.}
\newblock
\APACrefYearMonthDay{2021}{}{}.
\newblock
\begin{APACrefURL}
  [{2021-12-02}]\url{https://atmos.nmsu.edu/data_and_services/atmospheres_data/MARS/mcs.html}
  \end{APACrefURL}
\PrintBackRefs{\CurrentBib}

\bibitem [\protect \citeauthoryear {%
Medvedev%
\ \BBA {} Hartogh%
}{%
Medvedev%
\ \BBA {} Hartogh%
}{%
{\protect \APACyear {2007}}%
}]{%
medvedev2007winter}
\APACinsertmetastar {%
medvedev2007winter}%
\begin{APACrefauthors}%
Medvedev, A\BPBI S.%
\BCBT {}\ \BBA {} Hartogh, P.%
\end{APACrefauthors}%
\unskip\
\newblock
\APACrefYearMonthDay{2007}{}{}.
\newblock
{\BBOQ}\APACrefatitle {{Winter polar warmings and the meridional transport on
  Mars simulated with a general circulation model}} {{Winter polar warmings and
  the meridional transport on Mars simulated with a general circulation
  model}}.{\BBCQ}
\newblock
\APACjournalVolNumPages{Icarus}{186}{1}{97--110}.
\newblock
\begin{APACrefDOI} \doi{10.1016/j.icarus.2006.08.020} \end{APACrefDOI}
\PrintBackRefs{\CurrentBib}

\bibitem [\protect \citeauthoryear {%
Medvedev%
\ \protect \BOthers {.}}{%
Medvedev%
\ \protect \BOthers {.}}{%
{\protect \APACyear {2016}}%
}]{%
medvedev2016comparison}
\APACinsertmetastar {%
medvedev2016comparison}%
\begin{APACrefauthors}%
Medvedev, A\BPBI S.%
, Nakagawa, H.%
, Mockel, C.%
, Yiğit, E.%
, Kuroda, T.%
, Hartogh, P.%
\BDBL {}Jakosky, B\BPBI M.%
\end{APACrefauthors}%
\unskip\
\newblock
\APACrefYearMonthDay{2016}{}{}.
\newblock
{\BBOQ}\APACrefatitle {{Comparison of the Martian thermospheric density and
  temperature from IUVS/MAVEN data and general circulation modeling}}
  {{Comparison of the Martian thermospheric density and temperature from
  IUVS/MAVEN data and general circulation modeling}}.{\BBCQ}
\newblock
\APACjournalVolNumPages{Geophysical Research Letters}{43}{7}{3095--3104}.
\newblock
\begin{APACrefDOI} \doi{10.1002/2016GL068388} \end{APACrefDOI}
\PrintBackRefs{\CurrentBib}

\bibitem [\protect \citeauthoryear {%
Medvedev%
\ \BBA {} Yi{\u{g}}it%
}{%
Medvedev%
\ \BBA {} Yi{\u{g}}it%
}{%
{\protect \APACyear {2019}}%
}]{%
medvedev2019gravity}
\APACinsertmetastar {%
medvedev2019gravity}%
\begin{APACrefauthors}%
Medvedev, A\BPBI S.%
\BCBT {}\ \BBA {} Yi{\u{g}}it, E.%
\end{APACrefauthors}%
\unskip\
\newblock
\APACrefYearMonthDay{2019}{}{}.
\newblock
{\BBOQ}\APACrefatitle {Gravity waves in planetary atmospheres: Their effects
  and parameterization in global circulation models} {Gravity waves in
  planetary atmospheres: Their effects and parameterization in global
  circulation models}.{\BBCQ}
\newblock
\APACjournalVolNumPages{Atmosphere}{10}{9}{531}.
\PrintBackRefs{\CurrentBib}

\bibitem [\protect \citeauthoryear {%
Medvedev%
, Yi{\u{g}}it%
, Kuroda%
\BCBL {}\ \BBA {} Hartogh%
}{%
Medvedev%
\ \protect \BOthers {.}}{%
{\protect \APACyear {2013}}%
}]{%
medvedev2013general}
\APACinsertmetastar {%
medvedev2013general}%
\begin{APACrefauthors}%
Medvedev, A\BPBI S.%
, Yi{\u{g}}it, E.%
, Kuroda, T.%
\BCBL {}\ \BBA {} Hartogh, P.%
\end{APACrefauthors}%
\unskip\
\newblock
\APACrefYearMonthDay{2013}{}{}.
\newblock
{\BBOQ}\APACrefatitle {General circulation modeling of the {M}artian upper
  atmosphere during global dust storms} {General circulation modeling of the
  {M}artian upper atmosphere during global dust storms}.{\BBCQ}
\newblock
\APACjournalVolNumPages{Journal of Geophysical Research:
  Planets}{118}{10}{2234--2246}.
\PrintBackRefs{\CurrentBib}

\bibitem [\protect \citeauthoryear {%
Medvedev%
, Yiğit%
, Hartogh%
\BCBL {}\ \BBA {} Becker%
}{%
Medvedev%
\ \protect \BOthers {.}}{%
{\protect \APACyear {2011}}%
}]{%
medvedev2011influence}
\APACinsertmetastar {%
medvedev2011influence}%
\begin{APACrefauthors}%
Medvedev, A\BPBI S.%
, Yiğit, E.%
, Hartogh, P.%
\BCBL {}\ \BBA {} Becker, E.%
\end{APACrefauthors}%
\unskip\
\newblock
\APACrefYearMonthDay{2011}{}{}.
\newblock
{\BBOQ}\APACrefatitle {Influence of gravity waves on the {M}artian atmosphere:
  General circulation modeling} {Influence of gravity waves on the {M}artian
  atmosphere: General circulation modeling}.{\BBCQ}
\newblock
\APACjournalVolNumPages{Journal of Geophysical Research: Planets}{116}{E10}{}.
\newblock
\begin{APACrefDOI} \doi{10.1029/2011JE003848} \end{APACrefDOI}
\PrintBackRefs{\CurrentBib}

\bibitem [\protect \citeauthoryear {%
Montabone%
\ \protect \BOthers {.}}{%
Montabone%
\ \protect \BOthers {.}}{%
{\protect \APACyear {2020}}%
}]{%
montabone2020martian}
\APACinsertmetastar {%
montabone2020martian}%
\begin{APACrefauthors}%
Montabone, L.%
, Spiga, A.%
, Kass, D\BPBI M.%
, Kleinb{\"o}hl, A.%
, Forget, F.%
\BCBL {}\ \BBA {} Millour, E.%
\end{APACrefauthors}%
\unskip\
\newblock
\APACrefYearMonthDay{2020}{}{}.
\newblock
{\BBOQ}\APACrefatitle {Martian year 34 column dust climatology from {Mars
  Climate Sounder} observations: Reconstructed maps and model simulations}
  {Martian year 34 column dust climatology from {Mars Climate Sounder}
  observations: Reconstructed maps and model simulations}.{\BBCQ}
\newblock
\APACjournalVolNumPages{Journal of Geophysical Research:
  Planets}{125}{8}{e2019JE006111}.
\PrintBackRefs{\CurrentBib}

\bibitem [\protect \citeauthoryear {%
Nakagawa%
\ \protect \BOthers {.}}{%
Nakagawa%
\ \protect \BOthers {.}}{%
{\protect \APACyear {2020}}%
}]{%
Nakagawa_etal20b}
\APACinsertmetastar {%
Nakagawa_etal20b}%
\begin{APACrefauthors}%
Nakagawa, H.%
, Terada, N.%
, Jain, S\BPBI K.%
, Schneider, N\BPBI M.%
, Montmessin, F.%
, Yelle, R\BPBI V.%
\BDBL {}Jakosky, B\BPBI M.%
\end{APACrefauthors}%
\unskip\
\newblock
\APACrefYearMonthDay{2020}{}{}.
\newblock
{\BBOQ}\APACrefatitle {Vertical Propagation of Wave Perturbations in the Middle
  Atmosphere on {M}ars by {MAVEN/IUVS}} {Vertical propagation of wave
  perturbations in the middle atmosphere on {M}ars by {MAVEN/IUVS}}.{\BBCQ}
\newblock
\APACjournalVolNumPages{Journal of Geophysical Research:
  Planets}{125}{9}{e2020JE006481}.
\newblock
\begin{APACrefURL}
  \url{https://agupubs.onlinelibrary.wiley.com/doi/abs/10.1029/2020JE006481}
  \end{APACrefURL}
\newblock
\begin{APACrefDOI} \doi{10.1029/2020JE006481} \end{APACrefDOI}
\PrintBackRefs{\CurrentBib}

\bibitem [\protect \citeauthoryear {%
Neary%
\ \protect \BOthers {.}}{%
Neary%
\ \protect \BOthers {.}}{%
{\protect \APACyear {2020}}%
}]{%
neary2020explanation}
\APACinsertmetastar {%
neary2020explanation}%
\begin{APACrefauthors}%
Neary, L.%
, Daerden, F.%
, Aoki, S.%
, Whiteway, J.%
, Clancy, R\BPBI T.%
, Smith, M.%
\BDBL {}others%
\end{APACrefauthors}%
\unskip\
\newblock
\APACrefYearMonthDay{2020}{}{}.
\newblock
{\BBOQ}\APACrefatitle {Explanation for the increase in high-altitude water on
  {M}ars observed by {NOMAD} during the 2018 global dust storm} {Explanation
  for the increase in high-altitude water on {M}ars observed by {NOMAD} during
  the 2018 global dust storm}.{\BBCQ}
\newblock
\APACjournalVolNumPages{Geophysical Research Letters}{47}{7}{e2019GL084354}.
\newblock
\begin{APACrefDOI} \doi{10.1029/2019GL084354} \end{APACrefDOI}
\PrintBackRefs{\CurrentBib}

\bibitem [\protect \citeauthoryear {%
Parkinson%
\ \BBA {} Hunten%
}{%
Parkinson%
\ \BBA {} Hunten%
}{%
{\protect \APACyear {1972}}%
}]{%
parkinson1972spectroscopy}
\APACinsertmetastar {%
parkinson1972spectroscopy}%
\begin{APACrefauthors}%
Parkinson, T.%
\BCBT {}\ \BBA {} Hunten, D.%
\end{APACrefauthors}%
\unskip\
\newblock
\APACrefYearMonthDay{1972}{}{}.
\newblock
{\BBOQ}\APACrefatitle {Spectroscopy and acronomy of O2 on Mars} {Spectroscopy
  and acronomy of o2 on mars}.{\BBCQ}
\newblock
\APACjournalVolNumPages{Journal of the Atmospheric
  Sciences}{29}{7}{1380--1390}.
\PrintBackRefs{\CurrentBib}

\bibitem [\protect \citeauthoryear {%
Shaposhnikov%
, Medvedev%
, Rodin%
\BCBL {}\ \BBA {} Hartogh%
}{%
Shaposhnikov%
\ \protect \BOthers {.}}{%
{\protect \APACyear {2019}}%
}]{%
shaposhnikov2019seasonal}
\APACinsertmetastar {%
shaposhnikov2019seasonal}%
\begin{APACrefauthors}%
Shaposhnikov, D\BPBI S.%
, Medvedev, A\BPBI S.%
, Rodin, A\BPBI V.%
\BCBL {}\ \BBA {} Hartogh, P.%
\end{APACrefauthors}%
\unskip\
\newblock
\APACrefYearMonthDay{2019}{}{}.
\newblock
{\BBOQ}\APACrefatitle {Seasonal water “pump” in the atmosphere of {M}ars:
  Vertical transport to the thermosphere} {Seasonal water “pump” in the
  atmosphere of {M}ars: Vertical transport to the thermosphere}.{\BBCQ}
\newblock
\APACjournalVolNumPages{Geophysical Research Letters}{46}{8}{4161--4169}.
\PrintBackRefs{\CurrentBib}

\bibitem [\protect \citeauthoryear {%
Shaposhnikov%
, Medvedev%
, Rodin%
, Yiğit%
\BCBL {}\ \BBA {} Hartogh%
}{%
Shaposhnikov%
\ \protect \BOthers {.}}{%
{\protect \APACyear {2021}}%
}]{%
dmitry_s_shaposhnikov_2021_5749676}
\APACinsertmetastar {%
dmitry_s_shaposhnikov_2021_5749676}%
\begin{APACrefauthors}%
Shaposhnikov, D\BPBI S.%
, Medvedev, A\BPBI S.%
, Rodin, A\BPBI V.%
, Yiğit, E.%
\BCBL {}\ \BBA {} Hartogh, P.%
\end{APACrefauthors}%
\unskip\
\newblock
\APACrefYearMonthDay{2021}{}{}.
\newblock
\APACrefbtitle {{Martian Dust Storms and Gravity Waves: Disentangling Water
  Transport to the Upper Atmosphere}.} {{Martian Dust Storms and Gravity Waves:
  Disentangling Water Transport to the Upper Atmosphere}.}
\newblock
\APACaddressPublisher{}{Zenodo}.
\newblock
\begin{APACrefURL} \url{https://doi.org/10.5281/zenodo.5749676}
  \end{APACrefURL}
\newblock
\begin{APACrefDOI} \doi{10.5281/zenodo.5749676} \end{APACrefDOI}
\PrintBackRefs{\CurrentBib}

\bibitem [\protect \citeauthoryear {%
Shaposhnikov%
, Rodin%
\BCBL {}\ \BBA {} Medvedev%
}{%
Shaposhnikov%
\ \protect \BOthers {.}}{%
{\protect \APACyear {2016}}%
}]{%
shaposhnikov2016water}
\APACinsertmetastar {%
shaposhnikov2016water}%
\begin{APACrefauthors}%
Shaposhnikov, D\BPBI S.%
, Rodin, A\BPBI V.%
\BCBL {}\ \BBA {} Medvedev, A\BPBI S.%
\end{APACrefauthors}%
\unskip\
\newblock
\APACrefYearMonthDay{2016}{}{}.
\newblock
{\BBOQ}\APACrefatitle {{The water cycle in the general circulation model of the
  {M}artian atmosphere}} {{The water cycle in the general circulation model of
  the {M}artian atmosphere}}.{\BBCQ}
\newblock
\APACjournalVolNumPages{Solar System Research}{50}{2}{90--101}.
\newblock
\begin{APACrefURL} \url{http://link.springer.com/10.1134/S0038094616020039}
  \end{APACrefURL}
\newblock
\begin{APACrefDOI} \doi{10.1134/S0038094616020039} \end{APACrefDOI}
\PrintBackRefs{\CurrentBib}

\bibitem [\protect \citeauthoryear {%
Shaposhnikov%
\ \protect \BOthers {.}}{%
Shaposhnikov%
\ \protect \BOthers {.}}{%
{\protect \APACyear {2018}}%
}]{%
shaposhnikov2018modeling}
\APACinsertmetastar {%
shaposhnikov2018modeling}%
\begin{APACrefauthors}%
Shaposhnikov, D\BPBI S.%
, Rodin, A\BPBI V.%
, Medvedev, A\BPBI S.%
, Fedorova, A\BPBI A.%
, Kuroda, T.%
\BCBL {}\ \BBA {} Hartogh, P.%
\end{APACrefauthors}%
\unskip\
\newblock
\APACrefYearMonthDay{2018}{}{}.
\newblock
{\BBOQ}\APACrefatitle {Modeling the Hydrological Cycle in the Atmosphere of
  {M}ars: Influence of a Bimodal Size Distribution of Aerosol Nucleation
  Particles} {Modeling the hydrological cycle in the atmosphere of {M}ars:
  Influence of a bimodal size distribution of aerosol nucleation
  particles}.{\BBCQ}
\newblock
\APACjournalVolNumPages{Journal of Geophysical Research:
  Planets}{123}{2}{508--526}.
\PrintBackRefs{\CurrentBib}

\bibitem [\protect \citeauthoryear {%
Siddle%
, Mueller-Wodarg%
, Stone%
\BCBL {}\ \BBA {} Yelle%
}{%
Siddle%
\ \protect \BOthers {.}}{%
{\protect \APACyear {2019}}%
}]{%
Siddle_etal19}
\APACinsertmetastar {%
Siddle_etal19}%
\begin{APACrefauthors}%
Siddle, A.%
, Mueller-Wodarg, I.%
, Stone, S.%
\BCBL {}\ \BBA {} Yelle, R.%
\end{APACrefauthors}%
\unskip\
\newblock
\APACrefYearMonthDay{2019}{}{}.
\newblock
{\BBOQ}\APACrefatitle {Global characteristics of gravity waves in the upper
  atmosphere of {Mars} as measured by {MAVEN}/{NGIMS}} {Global characteristics
  of gravity waves in the upper atmosphere of {Mars} as measured by
  {MAVEN}/{NGIMS}}.{\BBCQ}
\newblock
\APACjournalVolNumPages{Icarus}{333}{}{12--21}.
\newblock
\begin{APACrefDOI} \doi{10.1016/j.icarus.2019.05.021} \end{APACrefDOI}
\PrintBackRefs{\CurrentBib}

\bibitem [\protect \citeauthoryear {%
Smith%
}{%
Smith%
}{%
{\protect \APACyear {2002}}%
}]{%
smith2002annual}
\APACinsertmetastar {%
smith2002annual}%
\begin{APACrefauthors}%
Smith, M\BPBI D.%
\end{APACrefauthors}%
\unskip\
\newblock
\APACrefYearMonthDay{2002}{}{}.
\newblock
{\BBOQ}\APACrefatitle {{The annual cycle of water vapor on Mars as observed by
  the Thermal Emission Spectrometer}} {{The annual cycle of water vapor on Mars
  as observed by the Thermal Emission Spectrometer}}.{\BBCQ}
\newblock
\APACjournalVolNumPages{Journal of Geophysical Research: Planets}{107}{E11}{}.
\newblock
\begin{APACrefURL} \url{http://doi.wiley.com/10.1029/2001JE001522}
  \end{APACrefURL}
\newblock
\begin{APACrefDOI} \doi{10.1029/2001JE001522} \end{APACrefDOI}
\PrintBackRefs{\CurrentBib}

\bibitem [\protect \citeauthoryear {%
Smith%
, Wolff%
, Clancy%
\BCBL {}\ \BBA {} Murchie%
}{%
Smith%
\ \protect \BOthers {.}}{%
{\protect \APACyear {2009}}%
}]{%
smith2009compact}
\APACinsertmetastar {%
smith2009compact}%
\begin{APACrefauthors}%
Smith, M\BPBI D.%
, Wolff, M\BPBI J.%
, Clancy, R\BPBI T.%
\BCBL {}\ \BBA {} Murchie, S\BPBI L.%
\end{APACrefauthors}%
\unskip\
\newblock
\APACrefYearMonthDay{2009}{}{}.
\newblock
{\BBOQ}\APACrefatitle {{Compact Reconnaissance Imaging Spectrometer
  observations of water vapor and carbon monoxide}} {{Compact Reconnaissance
  Imaging Spectrometer observations of water vapor and carbon
  monoxide}}.{\BBCQ}
\newblock
\APACjournalVolNumPages{Journal of Geophysical Research}{114}{E2}{E00D03}.
\newblock
\begin{APACrefURL} \url{http://doi.wiley.com/10.1029/2008JE003288}
  \end{APACrefURL}
\newblock
\begin{APACrefDOI} \doi{10.1029/2008JE003288} \end{APACrefDOI}
\PrintBackRefs{\CurrentBib}

\bibitem [\protect \citeauthoryear {%
Starichenko%
\ \protect \BOthers {.}}{%
Starichenko%
\ \protect \BOthers {.}}{%
{\protect \APACyear {2021}}%
}]{%
Starichenko_etal21}
\APACinsertmetastar {%
Starichenko_etal21}%
\begin{APACrefauthors}%
Starichenko, E\BPBI D.%
, Belyaev, D\BPBI A.%
, Medvedev, A\BPBI S.%
, Fedorova, A\BPBI A.%
, Korablev, O\BPBI I.%
, Trokhimovskiy, A.%
\BDBL {}Hartogh, P.%
\end{APACrefauthors}%
\unskip\
\newblock
\APACrefYearMonthDay{2021}{}{}.
\newblock
{\BBOQ}\APACrefatitle {Gravity Wave Activity in the {M}artian Atmosphere at
  Altitudes 20–160 km From {ACS/TGO} Occultation Measurements} {Gravity wave
  activity in the {M}artian atmosphere at altitudes 20–160 km from {ACS/TGO}
  occultation measurements}.{\BBCQ}
\newblock
\APACjournalVolNumPages{Journal of Geophysical Research:
  Planets}{126}{8}{e2021JE006899}.
\newblock
\begin{APACrefURL}
  \url{https://agupubs.onlinelibrary.wiley.com/doi/abs/10.1029/2021JE006899}
  \end{APACrefURL}
\newblock
\begin{APACrefDOI} \doi{https://doi.org/10.1029/2021JE006899} \end{APACrefDOI}
\PrintBackRefs{\CurrentBib}

\bibitem [\protect \citeauthoryear {%
Stone%
\ \protect \BOthers {.}}{%
Stone%
\ \protect \BOthers {.}}{%
{\protect \APACyear {2020}}%
}]{%
Stone_etal20}
\APACinsertmetastar {%
Stone_etal20}%
\begin{APACrefauthors}%
Stone, S\BPBI W.%
, Yelle, R\BPBI V.%
, Benna, M.%
, Lo, D\BPBI Y.%
, Elrod, M\BPBI K.%
\BCBL {}\ \BBA {} Mahaffy, P\BPBI R.%
\end{APACrefauthors}%
\unskip\
\newblock
\APACrefYearMonthDay{2020}{}{}.
\newblock
{\BBOQ}\APACrefatitle {Hydrogen escape from {M}ars is driven by seasonal and
  dust storm transport of water} {Hydrogen escape from {M}ars is driven by
  seasonal and dust storm transport of water}.{\BBCQ}
\newblock
\APACjournalVolNumPages{Science}{370}{6518}{824--831}.
\newblock
\begin{APACrefURL} \url{https://science.sciencemag.org/content/370/6518/824}
  \end{APACrefURL}
\newblock
\begin{APACrefDOI} \doi{10.1126/science.aba5229} \end{APACrefDOI}
\PrintBackRefs{\CurrentBib}

\bibitem [\protect \citeauthoryear {%
Vals%
\ \protect \BOthers {.}}{%
Vals%
\ \protect \BOthers {.}}{%
{\protect \APACyear {2019}}%
}]{%
vals2019study}
\APACinsertmetastar {%
vals2019study}%
\begin{APACrefauthors}%
Vals, M.%
, Spiga, A.%
, Forget, F.%
, Millour, E.%
, Montabone, L.%
\BCBL {}\ \BBA {} Lott, F.%
\end{APACrefauthors}%
\unskip\
\newblock
\APACrefYearMonthDay{2019}{}{}.
\newblock
{\BBOQ}\APACrefatitle {Study of gravity waves distribution and propagation in
  the thermosphere of {M}ars based on {MGS, ODY, MRO and MAVEN} density
  measurements} {Study of gravity waves distribution and propagation in the
  thermosphere of {M}ars based on {MGS, ODY, MRO and MAVEN} density
  measurements}.{\BBCQ}
\newblock
\APACjournalVolNumPages{Planetary and Space Science}{178}{}{104708}.
\PrintBackRefs{\CurrentBib}

\bibitem [\protect \citeauthoryear {%
Wang%
\ \BBA {} Richardson%
}{%
Wang%
\ \BBA {} Richardson%
}{%
{\protect \APACyear {2015}}%
}]{%
wang2015origin}
\APACinsertmetastar {%
wang2015origin}%
\begin{APACrefauthors}%
Wang, H.%
\BCBT {}\ \BBA {} Richardson, M\BPBI I.%
\end{APACrefauthors}%
\unskip\
\newblock
\APACrefYearMonthDay{2015}{}{}.
\newblock
{\BBOQ}\APACrefatitle {The origin, evolution, and trajectory of large dust
  storms on Mars during Mars years 24--30 (1999--2011)} {The origin, evolution,
  and trajectory of large dust storms on mars during mars years 24--30
  (1999--2011)}.{\BBCQ}
\newblock
\APACjournalVolNumPages{Icarus}{251}{}{112--127}.
\PrintBackRefs{\CurrentBib}

\bibitem [\protect \citeauthoryear {%
Yi{\u{g}}it%
\ \BBA {} Medvedev%
}{%
Yi{\u{g}}it%
\ \BBA {} Medvedev%
}{%
{\protect \APACyear {2015}}%
}]{%
yiugit2015internal}
\APACinsertmetastar {%
yiugit2015internal}%
\begin{APACrefauthors}%
Yi{\u{g}}it, E.%
\BCBT {}\ \BBA {} Medvedev, A\BPBI S.%
\end{APACrefauthors}%
\unskip\
\newblock
\APACrefYearMonthDay{2015}{}{}.
\newblock
{\BBOQ}\APACrefatitle {Internal wave coupling processes in {E}arth’s
  atmosphere} {Internal wave coupling processes in {E}arth’s
  atmosphere}.{\BBCQ}
\newblock
\APACjournalVolNumPages{Advances in Space Research}{55}{4}{983--1003}.
\PrintBackRefs{\CurrentBib}

\bibitem [\protect \citeauthoryear {%
Yi{\u{g}}it%
\ \BBA {} Medvedev%
}{%
Yi{\u{g}}it%
\ \BBA {} Medvedev%
}{%
{\protect \APACyear {2019}}%
}]{%
yiugit2019obscure}
\APACinsertmetastar {%
yiugit2019obscure}%
\begin{APACrefauthors}%
Yi{\u{g}}it, E.%
\BCBT {}\ \BBA {} Medvedev, A\BPBI S.%
\end{APACrefauthors}%
\unskip\
\newblock
\APACrefYearMonthDay{2019}{}{}.
\newblock
{\BBOQ}\APACrefatitle {Obscure waves in planetary atmospheres} {Obscure waves
  in planetary atmospheres}.{\BBCQ}
\newblock
\APACjournalVolNumPages{Physics Today}{72}{6}{}.
\PrintBackRefs{\CurrentBib}

\bibitem [\protect \citeauthoryear {%
Yi{\u{g}}it%
, Medvedev%
\BCBL {}\ \BBA {} Hartogh%
}{%
Yi{\u{g}}it%
\ \protect \BOthers {.}}{%
{\protect \APACyear {2018}}%
}]{%
yiugit2018influence}
\APACinsertmetastar {%
yiugit2018influence}%
\begin{APACrefauthors}%
Yi{\u{g}}it, E.%
, Medvedev, A\BPBI S.%
\BCBL {}\ \BBA {} Hartogh, P.%
\end{APACrefauthors}%
\unskip\
\newblock
\APACrefYearMonthDay{2018}{}{}.
\newblock
{\BBOQ}\APACrefatitle {Influence of gravity waves on the climatology of
  high-altitude {M}artian carbon dioxide ice clouds} {Influence of gravity
  waves on the climatology of high-altitude {M}artian carbon dioxide ice
  clouds}.{\BBCQ}
\newblock
\APACjournalVolNumPages{Annales Geophysicae}{36}{6}{1631--1646}.
\PrintBackRefs{\CurrentBib}

\bibitem [\protect \citeauthoryear {%
Yi\u{g}it%
, Aylward%
\BCBL {}\ \BBA {} Medvedev%
}{%
Yi\u{g}it%
\ \protect \BOthers {.}}{%
{\protect \APACyear {2008}}%
}]{%
Yigit_etal08}
\APACinsertmetastar {%
Yigit_etal08}%
\begin{APACrefauthors}%
Yi\u{g}it, E.%
, Aylward, A\BPBI D.%
\BCBL {}\ \BBA {} Medvedev, A\BPBI S.%
\end{APACrefauthors}%
\unskip\
\newblock
\APACrefYearMonthDay{2008}{}{}.
\newblock
{\BBOQ}\APACrefatitle {Parameterization of the effects of vertically
  propagating gravity waves for thermosphere general circulation models:
  Sensitivity study} {Parameterization of the effects of vertically propagating
  gravity waves for thermosphere general circulation models: Sensitivity
  study}.{\BBCQ}
\newblock
\APACjournalVolNumPages{Journal of Geophysical Research}{113}{}{}.
\newblock
\begin{APACrefDOI} \doi{10.1029/2008JD010135} \end{APACrefDOI}
\PrintBackRefs{\CurrentBib}

\bibitem [\protect \citeauthoryear {%
Yi\u{g}it%
, Medvedev%
, Benna%
\BCBL {}\ \BBA {} Jakosky%
}{%
Yi\u{g}it%
\ \protect \BOthers {.}}{%
{\protect \APACyear {2021b}}%
}]{%
Yigit_etal21b}
\APACinsertmetastar {%
Yigit_etal21b}%
\begin{APACrefauthors}%
Yi\u{g}it, E.%
, Medvedev, A\BPBI S.%
, Benna, M.%
\BCBL {}\ \BBA {} Jakosky, B.%
\end{APACrefauthors}%
\unskip\
\newblock
\APACrefYearMonthDay{2021b}{}{}.
\newblock
{\BBOQ}\APACrefatitle {Dust storm-enhanced gravity wave activity in the
  {M}artian thermosphere observed by {MAVEN} and implication for atmospheric
  escape} {Dust storm-enhanced gravity wave activity in the {M}artian
  thermosphere observed by {MAVEN} and implication for atmospheric
  escape}.{\BBCQ}
\newblock
\APACjournalVolNumPages{Geophysical Research Letters}{}{}{e2020GL092095}.
\newblock
\begin{APACrefDOI} \doi{10.1029/2020GL092095} \end{APACrefDOI}
\PrintBackRefs{\CurrentBib}

\bibitem [\protect \citeauthoryear {%
Yi\u{g}it%
, Medvedev%
\BCBL {}\ \BBA {} Hartogh%
}{%
Yi\u{g}it%
\ \protect \BOthers {.}}{%
{\protect \APACyear {2021a}}%
}]{%
Yigit_etal21a}
\APACinsertmetastar {%
Yigit_etal21a}%
\begin{APACrefauthors}%
Yi\u{g}it, E.%
, Medvedev, A\BPBI S.%
\BCBL {}\ \BBA {} Hartogh, P.%
\end{APACrefauthors}%
\unskip\
\newblock
\APACrefYearMonthDay{2021a}{}{}.
\newblock
{\BBOQ}\APACrefatitle {Variations of the {M}artian Thermospheric Gravity-wave
  Activity during the Recent Solar Minimum as Observed by {MAVEN}} {Variations
  of the {M}artian thermospheric gravity-wave activity during the recent solar
  minimum as observed by {MAVEN}}.{\BBCQ}
\newblock
\APACjournalVolNumPages{The Astrophysical Journal}{290}{2}{69}.
\newblock
\begin{APACrefURL} \url{https://doi.org/10.3847/1538-4357/ac15fc}
  \end{APACrefURL}
\newblock
\begin{APACrefDOI} \doi{10.3847/1538-4357/ac15fc} \end{APACrefDOI}
\PrintBackRefs{\CurrentBib}

\end{thebibliography}
	
	
	
	\newpage
	
	\begin{figure}[ht]
		\setcounter{figure}{0}
		\centerline{\includegraphics[width=40pc]{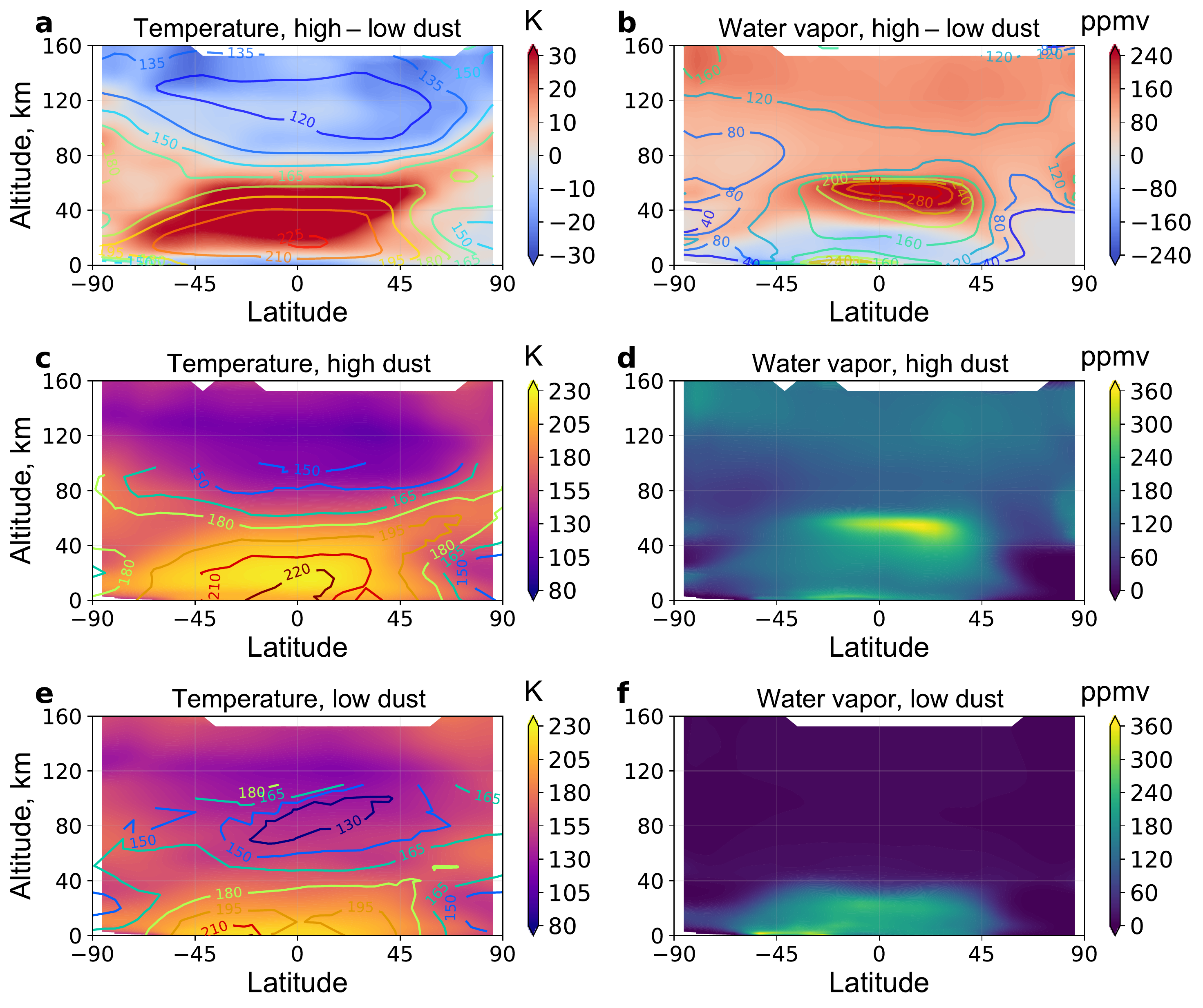}}
		\caption{Latitude-altitude cross-sections of temperature and water vapor at the beginning of the MY34 dust storm ($L_s=194^\circ - 200^\circ$).
			(\textbf{a}, \textbf{b}) Temperature (\textbf{a}) and water vapor (\textbf{b}) abundances simulated for the MY34 dust scenario (contours) and differences between the MY34 and MY28 dust scenarios (shaded).
			(\textbf{c}, \textbf{d}) Temperature (\textbf{c}) and water vapor (\textbf{d}) abundances simulated for the MY34 dust scenario (shaded) and temperature measurements derived from the Mars Climate Sounder (\textbf{c}, contours) \cite{mcs2021}.
			(\textbf{e}, \textbf{f}) Temperature (\textbf{e}) and water vapor (\textbf{f}) abundances simulated for the MY28 dust scenario (shaded) and temperature measurements derived from the Mars Climate Sounder (\textbf{e}, contours).
			The simulation for MY28 corresponds to the low-dust conditions, since the major storm has not started yet (see Figure~\ref{fig:annual}).}
		\label{fig:storm}
	\end{figure}
	
	\begin{figure}[ht]
		\setcounter{figure}{1}
		\centerline{\includegraphics[width=40pc]{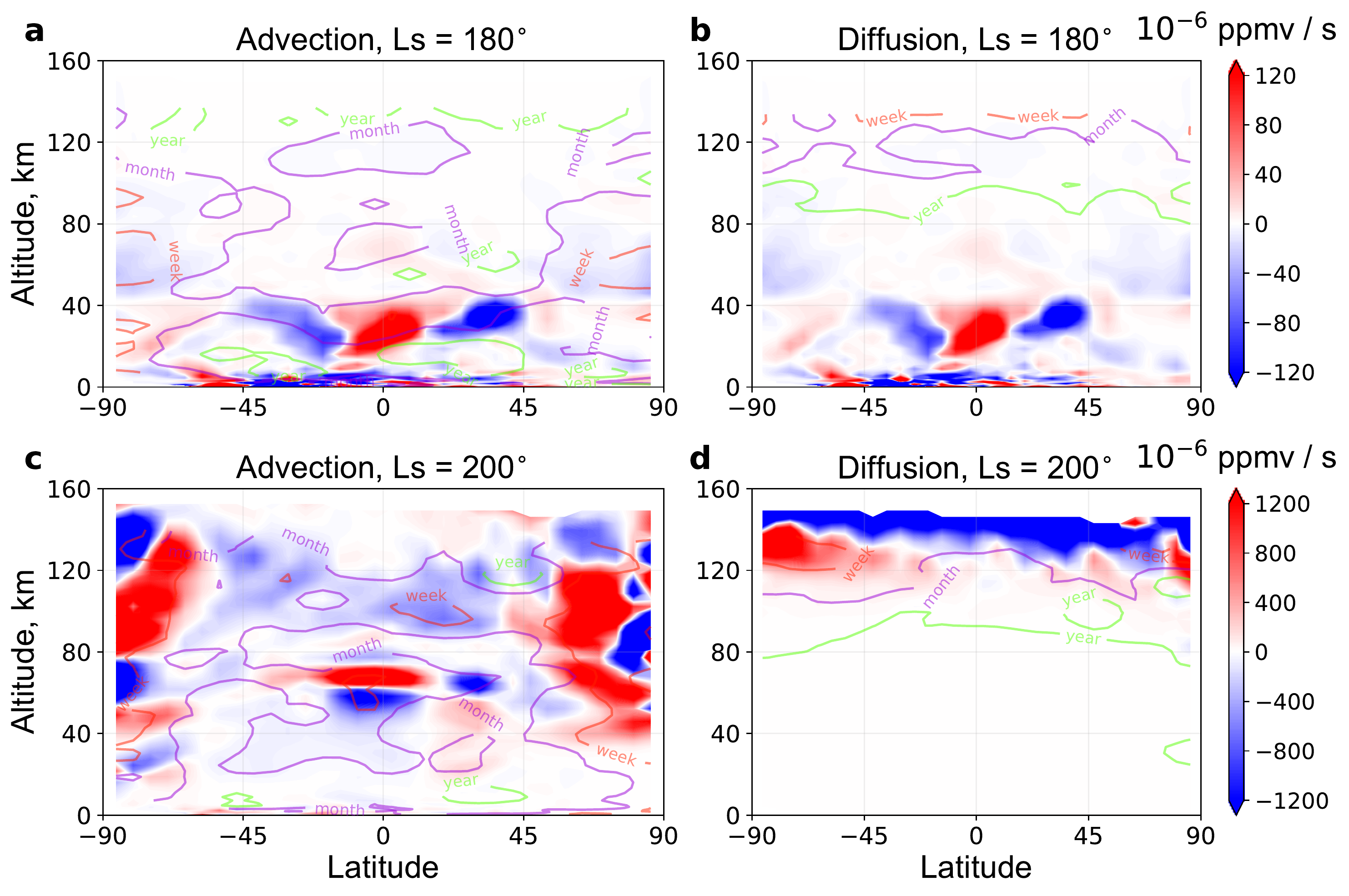}}
		\caption{The rate of change of water mixing ratio $\partial q/\partial t$ due to advection (left column) and molecular diffusion (right column) based on the corresponding terms from the continuity equation. The upper row is for a day before the onset of the MY34 dust storm, and the lower one is for a day in its midst. The contours show the characteristic time scales for advection and diffusion. The units refer to Martian week/month/year. \add{Note, that figures~\textbf{a}, \textbf{b} and \textbf{c}, \textbf{d} have different color scale.}}
		\label{fig:diffusion}
	\end{figure}
	
	\begin{figure}[ht]
		\setcounter{figure}{2}
		\centerline{\includegraphics[width=40pc]{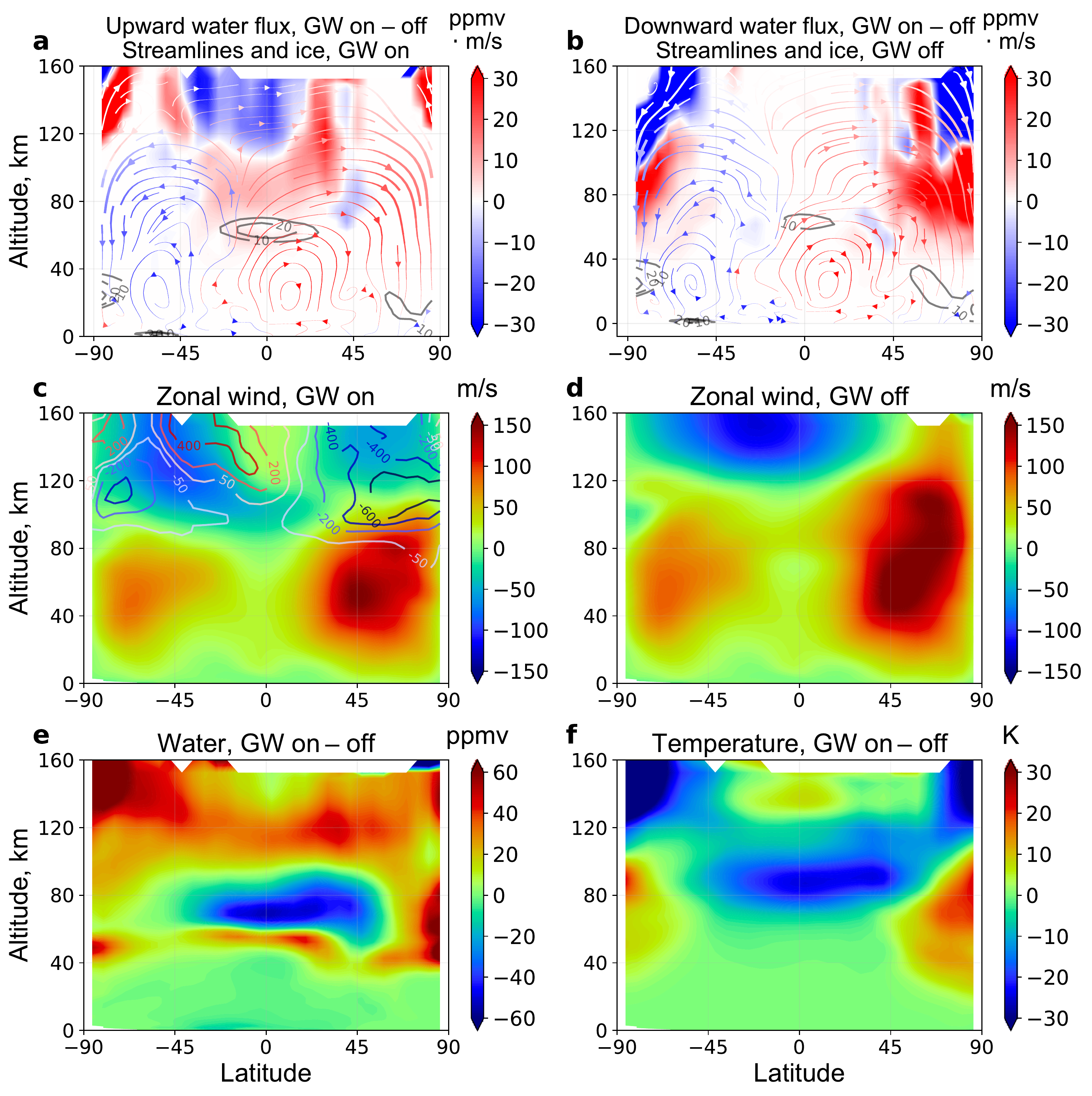}}
		\caption{Latitude-altitude cross-sections of the quantities simulated with the MY34 dust scenario at the beginning of the storm ($L_s$=$194^\circ$--$200^\circ$).
			(\textbf{a} and \textbf{b}, color shades): Differences between the runs with (``GW on") and without (``GW off") subgrid-scale gravity waves for the simulated upward (\textbf{a}) and downward (\textbf{b}) water vapor fluxes.
			(\textbf{a} and \textbf{b}, gray contours): Water ice (in ppmv) for the ``GW on" and ``GW off" runs, correspondingly.
			(\textbf{a} and \textbf{b}, streamlines with arrows): The meridional stream function for the ``GW on" (\textbf{a}) and ``GW off" (\textbf{b}) simulations.
			(\textbf{c}-\textbf{d}): Mean zonal wind (m~s$^{-1}$, shaded) and GW drag (m~s$^{-1}$~sol$^{-1}$, contours) for the ``GW on" and ``GW off" scenarios, correspondingly.
			Differences between water vapor VMR (\textbf{e}) and temperature (\textbf{f}) simulated in the ``GW on" and ``GW off" runs. All fields are averaged zonally and over 10 sols. }
		\label{fig:perihelion}
	\end{figure}
	
	\begin{figure}[ht]
		\setcounter{figure}{3}
		\centerline{\includegraphics[width=40pc]{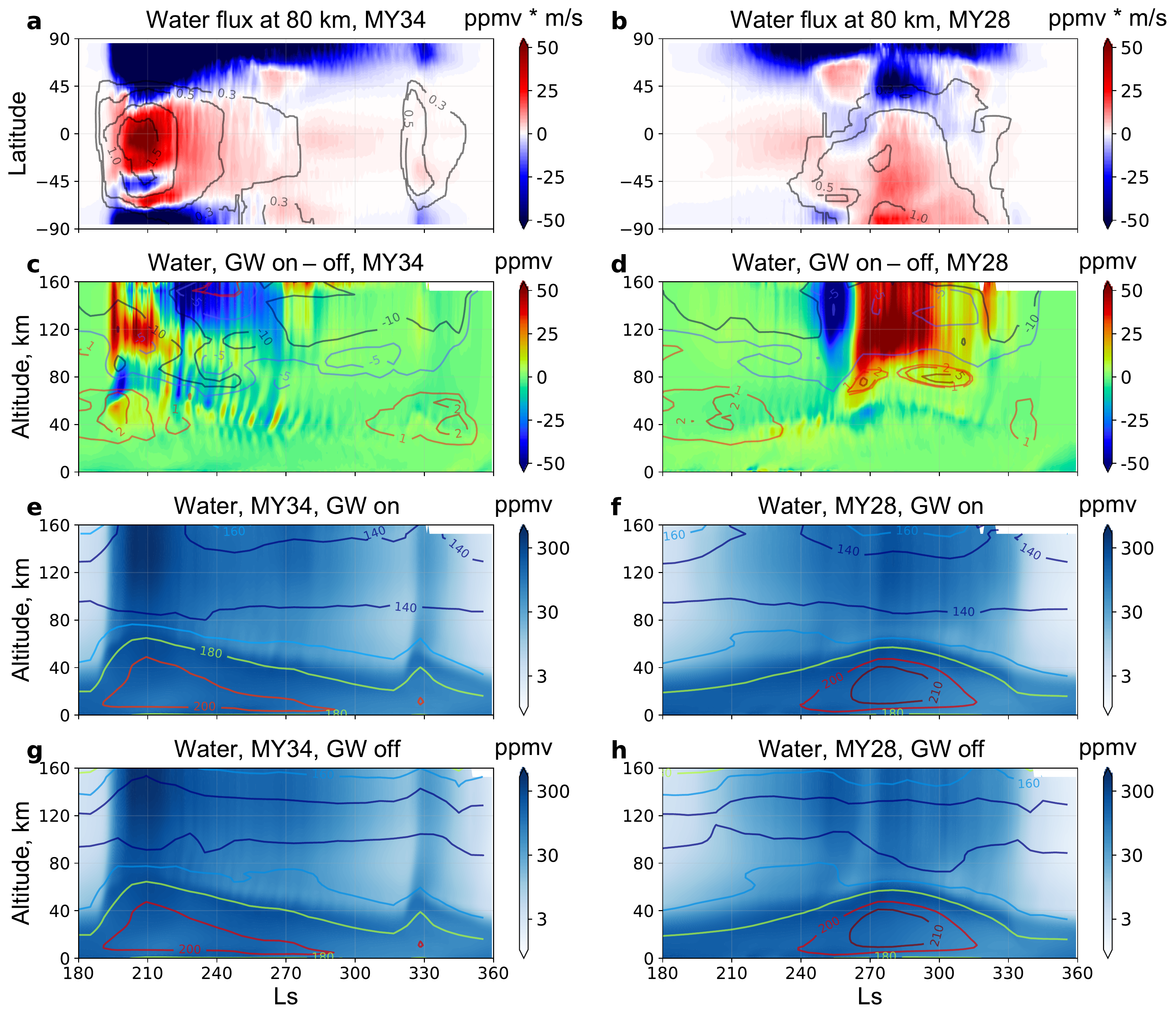}}
		\caption{(\textbf{a}, \textbf{b}) Latitude-seasonal cross-sections of the zonally averaged vertical water vapor flux at 80 km (shaded) simulated with gravity waves (GW) included and the IR optical dust opacity (black contours). (\textbf{c} - \textbf{h}) Altitude-seasonal cross-sections of the globally averaged differences (\textbf{c}, \textbf{d}) of water vapor (shaded) and temperature (contours) between the simulations with (\textbf{e}, \textbf{f}) and without (\textbf{g}, \textbf{h}) GWs. Left (\textbf{a}, \textbf{c}, \textbf{e}, \textbf{g}) and right (\textbf{b}, \textbf{d}, \textbf{f}, \textbf{h}) columns are for the MY34 and MY28 dust scenarios, correspondingly (see section~\ref{sec:MGCM}).}
		\label{fig:annual}
	\end{figure}
	
	\begin{figure}[ht]
		\setcounter{figure}{4}
		\centerline{\includegraphics[width=40pc]{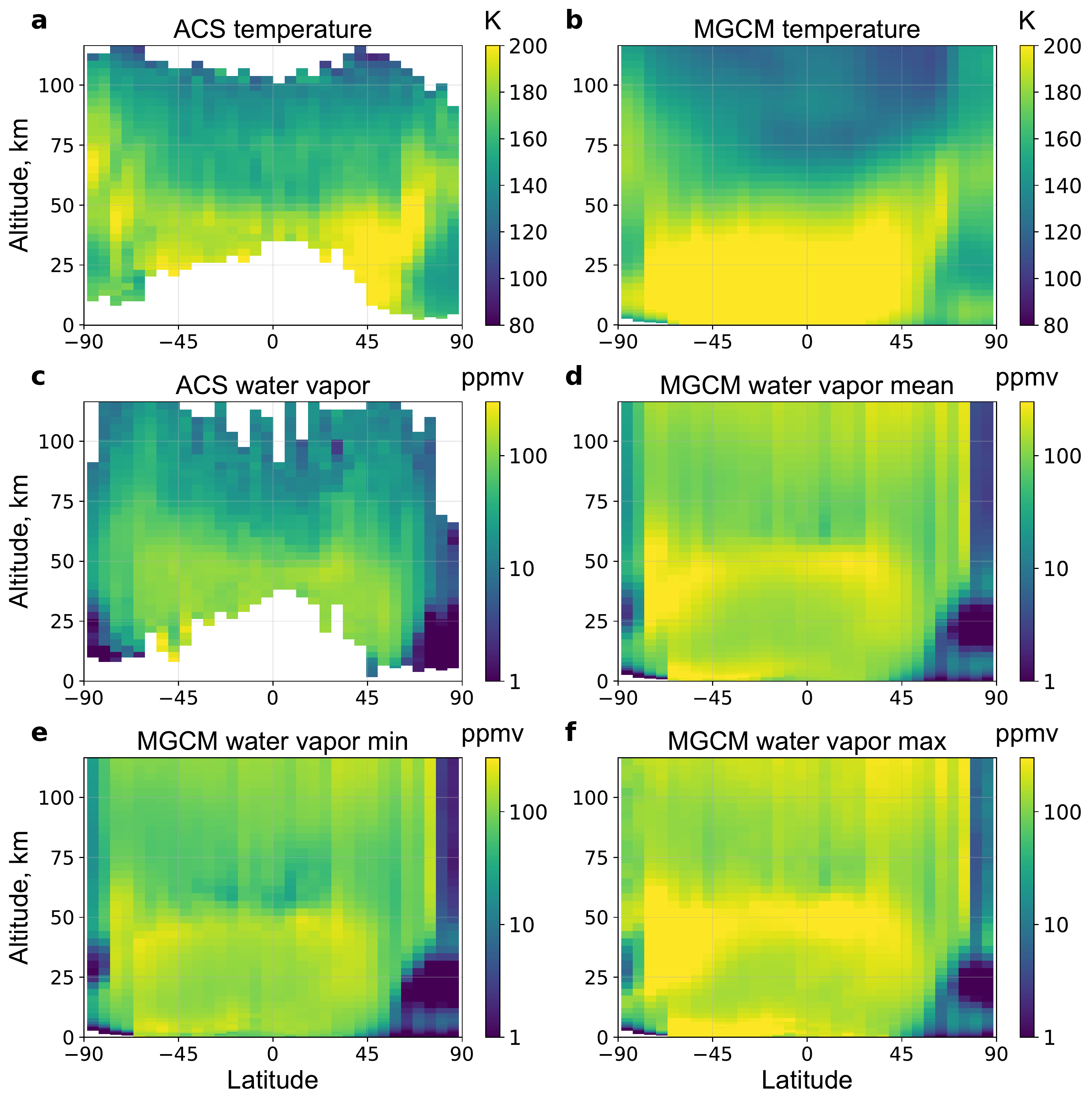}}
		\caption{Latitude-altitude distributions of temperature (\textbf{a}) and water vapor (\textbf{c}) derived from the Atmospheric Chemistry Suite (ACS) measurements \cite{fedorova2020stormy, Belyaev_etal21}. The corresponding distributions derived from simulations for the MY34 dust scenario (\textbf{b}, \textbf{d}, \textbf{e}, \textbf{f}). The ACS data and the model output are averaged over available measurements between $L_s=185^\circ - 267^\circ$. The MGCM data are taken at the same coordinates and local times as the measurements. The simulated water vapor composed of diurnal minimums (\textbf{e}) and maximums (\textbf{f}) at the same locations as the ACS data.}
		\label{fig:acs}
	\end{figure}
	
\end{document}